\documentclass[%
 reprint,
 amsmath,amssymb,
 aps,
 prf,
 twocolumn,
 superscriptaddress,
]{revtex4-2}

\usepackage{graphicx}
\usepackage{dcolumn}
\usepackage{bm}

\usepackage{xcolor}
\usepackage{float}

\begin{document}


\title{Triboelectrically mediated self-assembly and manipulation of drops at an interface}

 \author{Paul R. Kaneelil}
    \affiliation{Department of Mechanical and Aerospace Engineering, Princeton University, Princeton, New Jersey 08544, USA}
 \author{J. Pedro de Souza}
    \affiliation{Omenn-Darling Bioengineering Institute, Princeton University, Princeton, New Jersey 08544, USA}
 \author{Günther Turk}
    \affiliation{Princeton Materials Institute, Princeton University, Princeton, New Jersey 08544, USA}
    
 \author{Amir A. Pahlavan}
    \affiliation{Department of Mechanical Engineering and Materials Science, Yale University, New Haven, Connecticut 06511, USA}
 \author{Howard A. Stone}
    \affiliation{Department of Mechanical and Aerospace Engineering, Princeton University, Princeton, New Jersey 08544, USA}

\date{\today}

\begin{abstract}
The fluid-fluid interface is a complex environment for a floating object where the statics and dynamics may be governed by capillarity, gravity, inertia, and other external body forces. Yet, the alignment of these forces in intricate ways might result in beautiful pattern formation and self-assembly of these objects, as in the case of bubble rafts or colloidal particles. While interfacial self-assembly has been explored widely, controlled manipulation of floating objects, e.g. drops, at the fluid-fluid interface still remains a challenge largely unexplored. In this work, we reveal the self-assembly and manipulation of water drops floating at an oil-air interface. We show that the assembly occurs due to electrostatic interactions between the drops and their environment.  We highlight the role of the boundary surrounding the system by showing that even drops with a net zero electric charge can self-assemble under certain conditions. Using experiments and theory, we show that the depth of the oil bath plays an important role in setting the distance between the self-assembled drops. Furthermore, we demonstrate ways to manipulate the drops actively and passively at the interface.

\end{abstract}

\maketitle

\section{Introduction} \label{Ch4intro}
The ability to control and manipulate the classical elements -- air, water, earth, fire -- has always fascinated the human mind \cite{konietzko2013avatar}. The fascination has also influenced science and engineering; we focus on the topic of drop manipulation, in particular, which has been the subject of numerous studies. Manipulation of droplets and their control is important for many applications including ink-jet printing \cite{kuhn1979ink,basaran2013nonstandard}, biological testing \cite{barbulovic2008digital,hajji2020droplet}, cell-based screening \cite{whitesides2001soft}, and chemical reactions \cite{sun2020droplet,li2020droplet}. Among the oldest methods is dielectrophoresis, the motion of uncharged dielectric objects under the application of a non-uniform electric field, which has been used in many contexts to manipulate particles and drops \cite{batchelder1983dielectrophoretic,jones2002relationship}, and has even been used to design microfluidic chips that allow for the programmable motion of drops on a surface \cite{gascoyne2004dielectrophoresis,hunt2008integrated}. Electrowetting \cite{beni1981electro} and electrowetting-on-dielectric \cite{vallet1996electrowetting} are two other strategies that are commonly employed in drop manipulation research and applications \cite{pollack2000electrowetting,lee2002electrowetting,nelson2012droplet,peng2014ewod,li2019ionic}. Most of the other existing methods focus on manipulating drops on solid substrates using external fields: electric \cite{jin2022electrostatic,xu2022triboelectric,jin2023charge}, magnetic \cite{zhang2022reconfigurable,xu2023magnetocontrollable,liu2022simple}, ultrasonic \cite{yuan2023ultrasonic, luo2023contactless}, and light \cite{wang2022light,dong2023laser}. The success of these techniques often relies on the surfaces being superhydrophobic. 

While manipulation of drops on liquid surfaces rather than on solid ones offers an exciting alternative, it is challenging due to the dominating capillary forces and the so-called ``Cheerios" interactions \cite{gifford1971attraction,vella2005cheerios}. Previous works have, however, shown the self-assembly and pattern formation of drops at interfaces using vibration of a liquid bath \cite{couder2005walking,protiere2006particle,eddi2008wave, couchman2020free} and drop evaporation \cite{majhy2020attraction, li2023oil}. Yet, manipulating the drops or even their self-assembled patterns at the interface remain intractable.

In this article, we present an experimental system with water drops floating at an oil-air interface, where the drops can be manually controlled and/or autonomously self-assembled in ways that depend on the boundary conditions. Furthermore, we show that the self-assembly is  programmable, adding a layer of controllability. The two configurations that we present to manipulate drops at interfaces both rely on electrostatic interactions. In the first configuration, we show that the static charges on nearby surfaces can polarize neutrally charged drops and then be used to pattern or move the drops along the interface. The second configuration explores the assembly of charged drops at the oil-air interface. We identify the height of the oil bath as a key parameter that can be used to control the distance between the drops. Making an analogy with a classic electrostatics problem and using the method of images, we show that the height of the bottom boundary can be patterned to further control the location of the drops.

\section{Experimental setup}

\begin{figure*}[t]
\begin{center}
\includegraphics[width=0.7\textwidth]{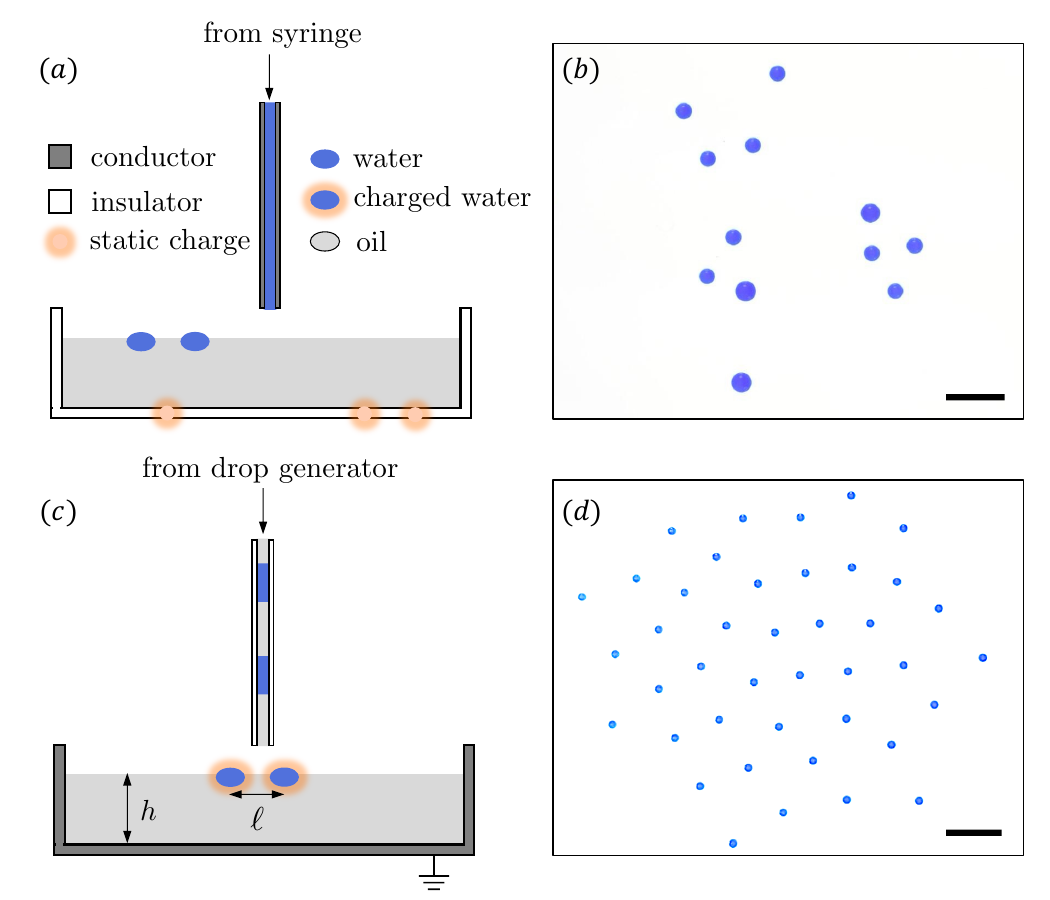}
\caption[The experimental setup for the two configurations.]{\label{ch4_schematics} The experimental setup for the two configurations. (a) Schematic of the first configuration where the drops are dispensed from a syringe through a metal needle. The drops, which have a neutral charge, float at the oil-air interface and the container is an insulator (dielectric). The yellow spots on the container schematically show localized static charge that might be present on the dielectric. (b) Top view experimental image showing the preferential self-assembly of the drops in configuration 1. (c) Schematic of the second configuration where the drops are produced by a microfluidic drop generator and are dispensed onto the container through an insulating tube. The drops in this configuration have a net positive electric charge and the container is a conductor that is grounded. (d) Top view experimental image showing the self-assembly of the charged drops. Scale bars in (b) and (d) represent 1.0 cm. }
\end{center}
\end{figure*}

The two experimental configurations are summarized in Fig. \ref{ch4_schematics}: panels (a) and (b) correspond to the first configuration, and panels (c) and (d) correspond to the second. Both cases involve water drops floating at an oil-air interface. We use Fluorinert FC-40 (Sigma-Aldrich) as the oil phase with a relative permittivity $\epsilon_r=1.9$, surface tension $\gamma_o = 17.3 \pm 0.3$ mN/m, viscosity $\mu_o = 4.01$ mPa.s, and density $\rho_o = 1855$ kg/m$^3$. The permittivity and density are crucial properties for all observations presented here: a low relative permittivity allows for electrostatic interactions through the oil without screening and a high density allows for the water drops to float. Experiments with Novec 7500 (3M) as the oil phase, $\epsilon_r = 5.8$, did not show self-assembly. The aqueous phase is deionized water dyed with Erioglaucine disodium salt (0.4 wt.\%; Sigma-Aldrich). 

In the first experimental configuration, the water drops are not charged and are dispensed onto an oil-filled petri dish that is an insulator (dielectric). We show that the uncharged drops can still be polarized by static charges on the surface of the container, or any nearby objects, and therefore can be assembled and controlled at the oil-air interface. As shown in Fig. \ref{ch4_schematics}(a), water is dispensed continuously from a syringe through a conductor (metal needle) to ensure neutral charge of the drops deposited onto the oil bath. Figure \ref{ch4_schematics}(b) shows a top view of the drops self-assembled at the interface in configuration 1. 

In the second configuration, the drops are electrically charged and are dispensed onto a petri dish covered with aluminum foil and electrically grounded, as shown in Fig. \ref{ch4_schematics}(c). The drops are generated with a microfluidic T-junction device and flow through a teflon tube (length L = 20 cm and inner diameter 0.38 mm), where they pick up a net positive charge due to the charge separation that occurs in the electrical double layer \cite{mccarty2008electrostatic, sun2015using, choi2013spontaneous}. Figure \ref{ch4_schematics}(d) shows the self-assembly of the charged drops, which exhibit more order than the polarized case (Fig. \ref{ch4_schematics}(a,b)).

\section{Polarized drops}

\begin{figure}[bt]
\begin{center}
\includegraphics[width=0.5\textwidth]{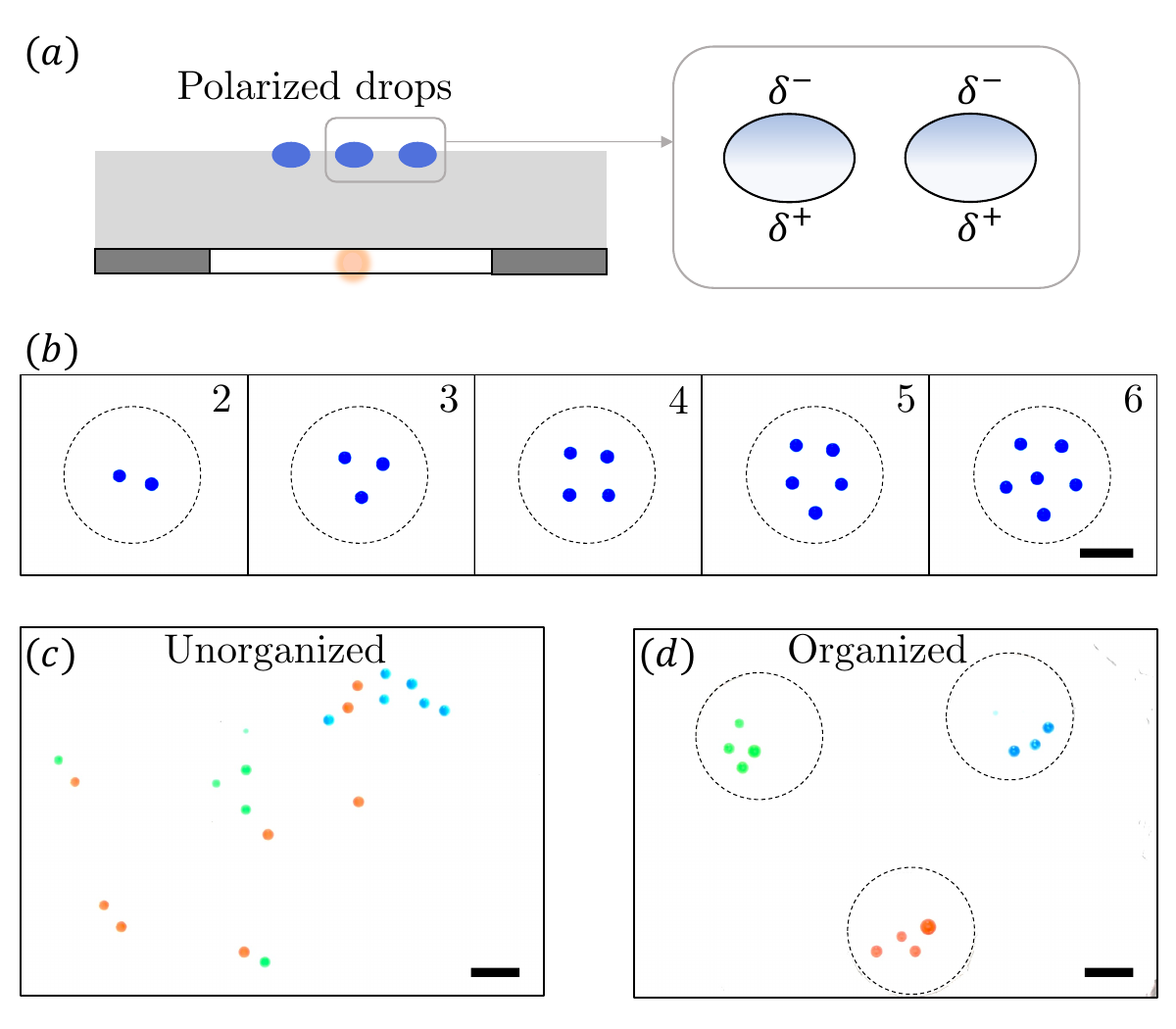}
\caption[The preferential assembly of polarized but uncharged drops.]{\label{ch4_insulatingbc} The preferential assembly of polarized but uncharged drops (configuration 1). (a) Schematic showing the experimental setup where the container is covered with aluminum foil (dark gray) with only a small window in the middle cut out to expose the dielectric material. The inset shows a schematic of the drops polarized, in this case by a negative charge on the surface below. (b) Experimental images showing drops assembled near the static charge that are present on the dielectric surface. Note that the interior of the dotted circle represents the area that is exposed to the dielectric material underneath. The assembled drops from $N=2-6$ show 2D crystalline order. (c) Experimental image showing an ``unorganized" state where the entire dielectric surface of the container is exposed. The drops dyed with different colors are randomly dispensed over the surface and get attracted by nearby charges. (d) Experimental image showing an ``organized" state, where three windows expose the dielectric surface of the container. Drops are dispensed inside each window based on their color such that each window contains a set of different colored drops. Scale bars in (b-d) represent 1.0 cm.}
\end{center}
\end{figure}

While we rely on electrostatic interactions to manipulate and control the drops at the interface, we show that it is not necessary for the drops to have a net electric charge. Figure \ref{ch4_insulatingbc}(a) shows a schematic of the oil-water system in a petri dish covered with aluminum foil with a small circular window cut out in the middle. The window exposes the plastic underneath, which is a dielectric surface prone to static charges. In fact, we can place charges on dielectric surfaces by bringing two dielectric surfaces into contact, causing a charge separation that leaves one surface with an excess negative, and the other with an excess positive charge \cite{mccarty2008electrostatic}. In our experiments, we do this by touching the exposed surface of the petri dish with a gloved finger after neutralizing the surface with a Zerostat pistol (Sigma-Aldrich).

Water is a strong dielectric ($\epsilon_r=78.5$) and can be  polarized easily in the presence of an electric field. When the uncharged water drops encounter the field produced by the static charges and become polarized, as shown in the inset in Fig. \ref{ch4_insulatingbc}(a), there are two consequences: the drops are attracted toward the charge causing the drops to move to the location on the interface above the charge, which then leads to dipole-dipole repulsion between the drops. The combination of these two effects results in self-assembly of the drops at the interface above the location of the static charges. Figure \ref{ch4_insulatingbc}(b) shows the patterns formed by $N$ number of drops as they assemble above the exposed window on the petri dish; the dotted circles in the images represent the outline of the exposed window. The reason for covering most of the petri dish with aluminum foil and only exposing a small window is to localize the effect and to eliminate unwanted field interactions. As shown in the figure, the drops show 2D crystalline order from $N=2$ to $N=6$. Adding more drops resulted in a breakdown of this order and coalescence between some of the drops. This breakdown could be because the attractive force bringing the drops toward the charge is stronger than the dipole-dipole repulsion between the drops. The capillary length of the system is $l_c=\sqrt{\gamma_o/(\rho_o g)} \approx 1$ mm, which means that an additional attractive force contribution from the Cheerios effect might also be present and dominant for separation distances close to $l_c$.

The polarization effect of the water drops can be further leveraged to sort and organize drops at the interface. Figure \ref{ch4_insulatingbc}(c) shows an ``unorganized" state of the drops, dyed red, blue, or green, when the petri dish below is fully exposed. The randomly distributed static charges on the boundary leads to a random distribution of the drops along the interface. To achieve a more ``organized" pattern, where the location of the drops can be dictated in advance, we covered the petri dish with aluminum foil only exposing three circular areas as shown in Fig. \ref{ch4_insulatingbc}(d). Each of the colored drops was released above its corresponding window. Thus, we were able to collect and retain the three different colored drops in three different regions. Patterning the surface of the container with static charges is therefore a simple way to dictate the location of the uncharged floating drops and to create local self-assembled colonies.

\section{Charged drops}
             
The focus so far has been on understanding the collective behavior of uncharged, polarized drops at the oil-air interface in an insulating container. Here, we switch our attention to the second configuration (Fig.~\ref{ch4_schematics}(c-d)): charged drops in a conductive container.

In the experiments, we dispense electrically charged drops onto the center of the container with the oil bath at height $h$. The drops have a radius of about $R=1.5$ mm and have a charge of approximately $Q_d=0.07 \pm 0.01 $ nC as measured using a Faraday cup and NanoCoulomb meter (Monroe Electronics; Model 284; see SI Fig. S1). As subsequent drops are added, the previous drops gets pushed away from the center towards the edges due to electrostatic repulsion and due to the flow of oil in the bath. Therefore, in order to ensure that all experiments start with a similar initial condition and that the patterns we observe are independent of the dynamics that occur during the introduction of the drops, upon stopping the drop dispensation we bring the drops closer together at the interface and constrict them using a dielectric stylus, before allowing them to spread apart and relax. 

\begin{figure*}[t]
\begin{center}
\includegraphics[width=0.95\textwidth]{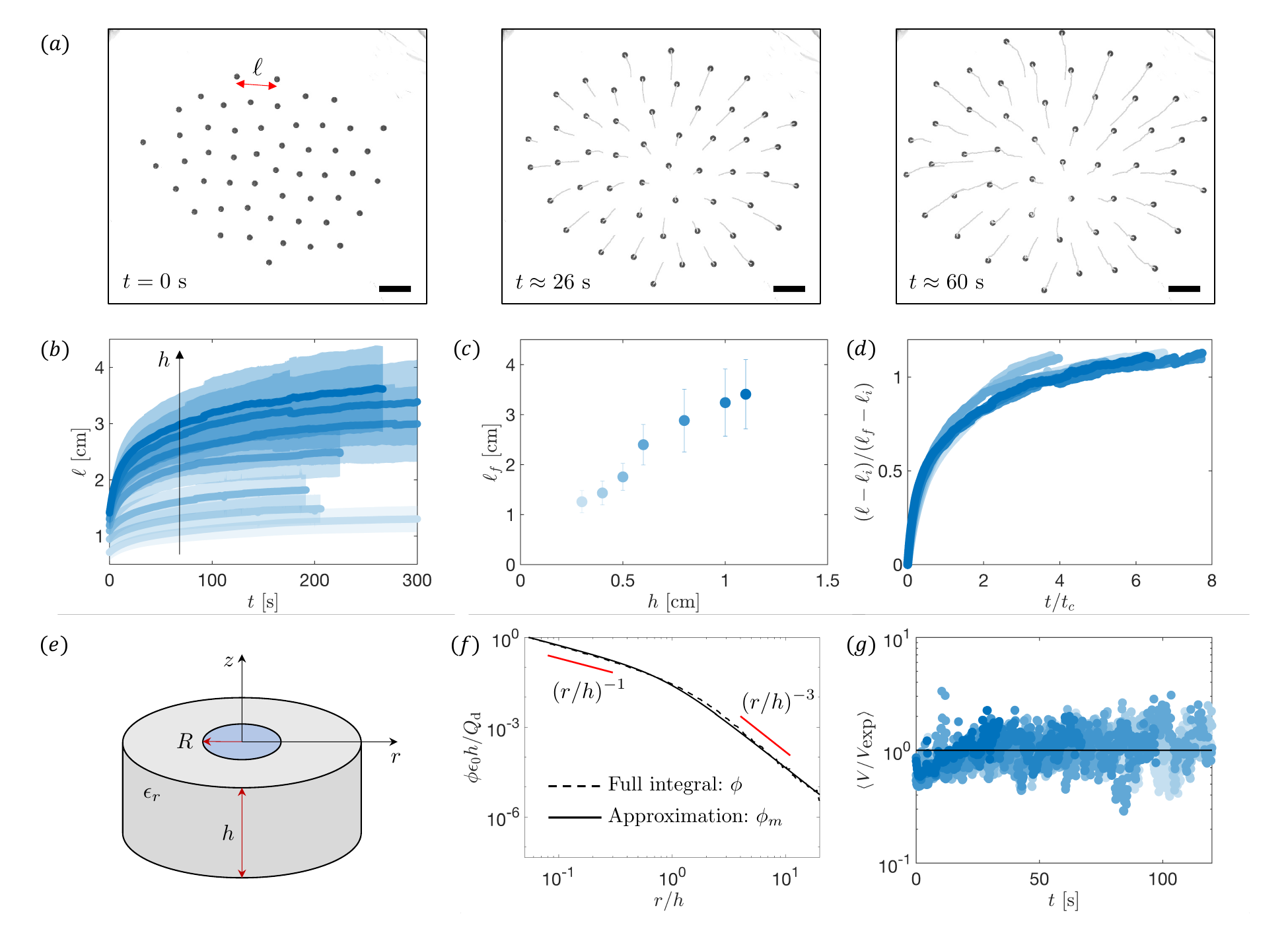}
\caption{\label{ch4_d-h} The distance between the drops as a function of the height $h$ of the oil bath. (a) Image sequence showing the self-assembly of drops for a case with $h=0.4$ cm at the moment the assembly was allowed to relax, $t=0$ s, and later at $t \approx 26$ s and $t \approx 60$ s. The gray tails map  the trajectory of each drop and the scale bars represent 1 cm. (b) The time evolution of the average distance $\ell$ between the drops for several $h$ values, as the drop assembly relaxes and reaches a quasi-equilibrium configuration with typical spacing $\ell_{f}$. From the lightest to the darkest, the marker colors correspond to $h=0.3,~0.4,~0.5,~0.6,~0.8,~1.0,$ and $1.1$ cm. The shaded error bars are standard deviations from averaging the distance for all the drops in the frame. (c) The quasi-equilibrium distance $\ell_{f}$ as a function of $h$. (d) Rescaling the $x$ and $y$ axis, all curves can be collapsed. Here, $t_c=(\ell_{f}-\ell_i)/V_0$ and $V_0$ is the initial velocity of the drops. (e) Schematic showing the geometry considered in the theoretical model. The drop is represented as a flat circle of radius $R$, which is a distance $h$ from the bottom conductive boundary. (f) The theoretical result showing the exact electrostatic potential $\phi$ (dashed line) and the approximate solution $\phi_m$ (Eq. \ref{eqch4potential}; solid line) at the interface , $z=0$, in the point charge limit ($R\rightarrow \infty$). (g) The ratio of the magnitude of the theoretical velocity of a drop to the magnitude of the experimentally measured velocity averaged over all drops, $\langle V/V_{\textrm{exp}}\rangle$, is plotted as a function of time. The solid black line represents $\langle V/V_{\textrm{exp}}\rangle=1$.}
\end{center}
\end{figure*}

For a case with $h=0.4$ cm, Fig.~\ref{ch4_d-h}(a) shows an image sequence at the moment the array was allowed to relax, which we label $t=0$, and later at $t \approx 26$ s and $t \approx 60$ s. The gray tails map the trajectory of each drop from its initial position. Note that the length of the tails is indicative of the speed of the drops, which for reasons of symmetry is larger for the drops along the periphery. In what follows, we will show that this system is dominated by electrostatic repulsion. The drops are always moving away from each other although they will reach a quasi-equilibrium configuration within the observation time of the experiments.

\subsection{Effect of the height of the oil bath}

We analyzed how the distance between the drops evolves as each of the assemblies with varying $h$ reach their quasi-equilibrium pattern. The distance $\hat{\ell}$ for a given drop is defined as the average distance to its three closest neighbors. We chose to consider three neighbors since the drops on the outer edge of the pattern might only have three neighbors -- two on each side and one towards the interior. Figure \ref{ch4_d-h}(b) shows the average distance $\ell$, representing the average of $\hat{\ell}$ for all the drops on the surface, as a function of time $t$, with $t=0$ being defined as the time when the assemblies were allowed to relax from the initial state of confinement. We observe that $\ell$ increases over time and effectively approaches a constant value. To characterize this behavior, we define the quasi-equilibrium distance $\ell_f$ as the average of the distance when the local fractional change in $\ell$ is below a threshold: $\Delta \ell/\ell < 10^{-4}$. In Fig. \ref{ch4_d-h}(c), we plot $\ell_f$ as a function of $h$. As $h$ is increased, $\ell_f$ also increases. But for large $h$, $\ell_f$ is only weakly dependent on $h$, which is expected since the effect of the bottom boundary will decrease as the distance from the boundary is increased. In Fig. \ref{ch4_d-h}(d), we rescale the time evolution of $\ell$ by only considering the fractional change in distance $(\ell-\ell_i)/(\ell_f-\ell_i)$ over time, where $\ell_i=\ell|_{t=0}$. We rescale time with an empirical time scale $t_c = (\ell_f-\ell_i)/V_0$, where $V_0 = \frac{d\ell_i}{dt}$ is the average initial velocity of the drops as calculated from the experimental data. The collapse of the data indicates that the initial spacing $\ell_i$ of the drops and height $h$ of the bath set the time scale of the dynamics.


The dependence of the collective behavior of the drops on the bath height $h$ is captured by the representative electrostatic potential $\phi$ of the system and can be explained with a simplified model. Consider the geometry shown in Fig. \ref{ch4_d-h}(e), where the drop is assumed to be a flat circle of radius $R$, located at the origin, and having a uniform charge distribution. We consider a domain that is infinite in the radial direction $r$ and semi-infinite in the axial direction $z$, where there is a finite height $h$ of oil below the drop with a relative permittivity $\epsilon_r$ and an infinite layer of air above the drop. An exact analytical solution does not exist for $\phi$ in closed form. But, the approximate solution, labelled $\phi_m$, representing the electrostatic potential at the interface generated by one drop can be written as (see SI Sec. II)
\begin{equation}
    \phi_m(r,z=0)= \frac{Q_d h^2}{2 \pi \epsilon_0 r[(\epsilon_r+1)h^2+\epsilon_r^2r^2]},
    \label{eqch4potential}
\end{equation}
where $Q_d$ is the total charge of the drop, $\epsilon_0 = 8.854 \times 10^{-12}$ F/m is the permittivity of free space, and $r$ is the distance from the center of the drop. Our model captures the dielectric discontinuity at the oil-air interface and the conductive boundary condition at the bottom. Figure \ref{ch4_d-h}(f) shows $\phi_m$ plotted along with the exact numerical solution $\phi$ in the point charge limit ($R\rightarrow 0$), evaluated at $z=0$, in dimensionless form. Note that for shallow bath heights, corresponding to large $r/h$, the dimensionless potential scales as $(r/h)^{-3}$ which means that the potential has a quadratic dependence with the height, as also seen from Eq. \ref{eqch4potential}. A detailed calculation can be found in the Supplementary Information (see SI Fig. S2).


It is important to note that the patterns we observe in Fig. \ref{ch4_d-h}(a) and the separation distances $\ell_f$ that we measure are mesoscale, with $\ell_f \gg l_c$, which clearly indicates that the electrostatic interactions are much stronger than the Cheerios effect (See SI Sec. III and Fig. S3). We analyze the velocity of drops spreading as a function of time and compare with the theoretical prediction using the calculated electrostatic potential in order to validate that the system is dominated by electrostatic repulsion. Assuming that inertia is negligible, the evolution of the speed of a single drop for a purely repulsive system can be derived by balancing the viscous drag force with the electrostatic force. Taking the viscous drag to be that of a sphere in a bulk fluid modified by a fitting parameter $\beta$ (see SI Sec. IV and Fig. S4), which accounts for any excess drag due to the presence of the interface and the bottom surface of the container in shallow cases, the velocity is $\textbf{V} = \frac{1}{6 \beta \pi \mu_o R} \frac{dU_e}{dr}\hat{\textbf{e}}_r$. Here $U_e = Q_d ~\phi_m(r,z=0)$ is the electrostatic energy and $\hat{\textbf{e}}_r$ is the unit vector from the drop towards its neighbor. The theoretical velocity of a two-body system separated by a distance $r$ is then (see Supplementary Information)
\begin{equation}
    \textbf{V} = \frac{-Q_d^2 h^2 \big[(\epsilon_r+1)h^2 + 3\epsilon_r^2 r^2 \big]}{12 \beta \pi^2 \mu_o \epsilon_0 R r^2 \big[(\epsilon_r+1)h^2 + \epsilon_r^2 r^2 \big]^2} ~\hat{\textbf{e}}_r.
    \label{eqch4_Vfull}
\end{equation}

We calculate the theoretical velocity of each of the drops in the container at every time step by taking the distances and unit vectors as input from the experimental data and summing the two-body interactions between a drop and six of its closest neighbors to calculate the resultant velocity vector. Considering more than six neighbors had negligible effect on the resultant velocity vector. The magnitude $V$ of this theoretical velocity can be compared to the magnitude of the experimentally measured velocity $V_{\textrm{exp}}$ for each drop in the container. The ratio of the two velocities averaged over all the drops for a given time step, $\langle V/V_{\textrm{exp}}\rangle$, is plotted as a function of time in Fig. \ref{ch4_d-h}(g). We observe that $\langle V/V_{\textrm{exp}}\rangle$ cluster around unity, as depicted by the black line, which suggests that the model captures the major experimental features. The scatter may be attributed to the point charge approximation of the drop in the model, the nonlinear many-body interactions not considered in the model, and the extra viscous stresses introduced by the lubricating film underneath the drop for small bath heights. Nevertheless, the reasonable agreement between the experimental data and the model prediction suggests that the dynamics are dominated by electrostatic repulsion.

We further calculate the time that it would take for two drops starting at the center of the container to repel and reach the side walls of the container. The distance from the center to the side walls of the container is approximately $l = 7.5$ cm. The time $t_f$ that it would take the drops to traverse that distance can be calculated by $t_f = \int_0^l (1/V)~dr$, which is about 3 hours and 40 hours, respectively, for $h=1.1$ cm and $h=0.3$ cm. These large durations are a result of the quickly decaying strength of the electrostatic potential, which scales as $r^{-3}$, and is also the reason why the distance between the drops appears to approach a quasi-equilibrium value.

\begin{figure*}[t]
\begin{center}
\includegraphics[width=0.78\textwidth]{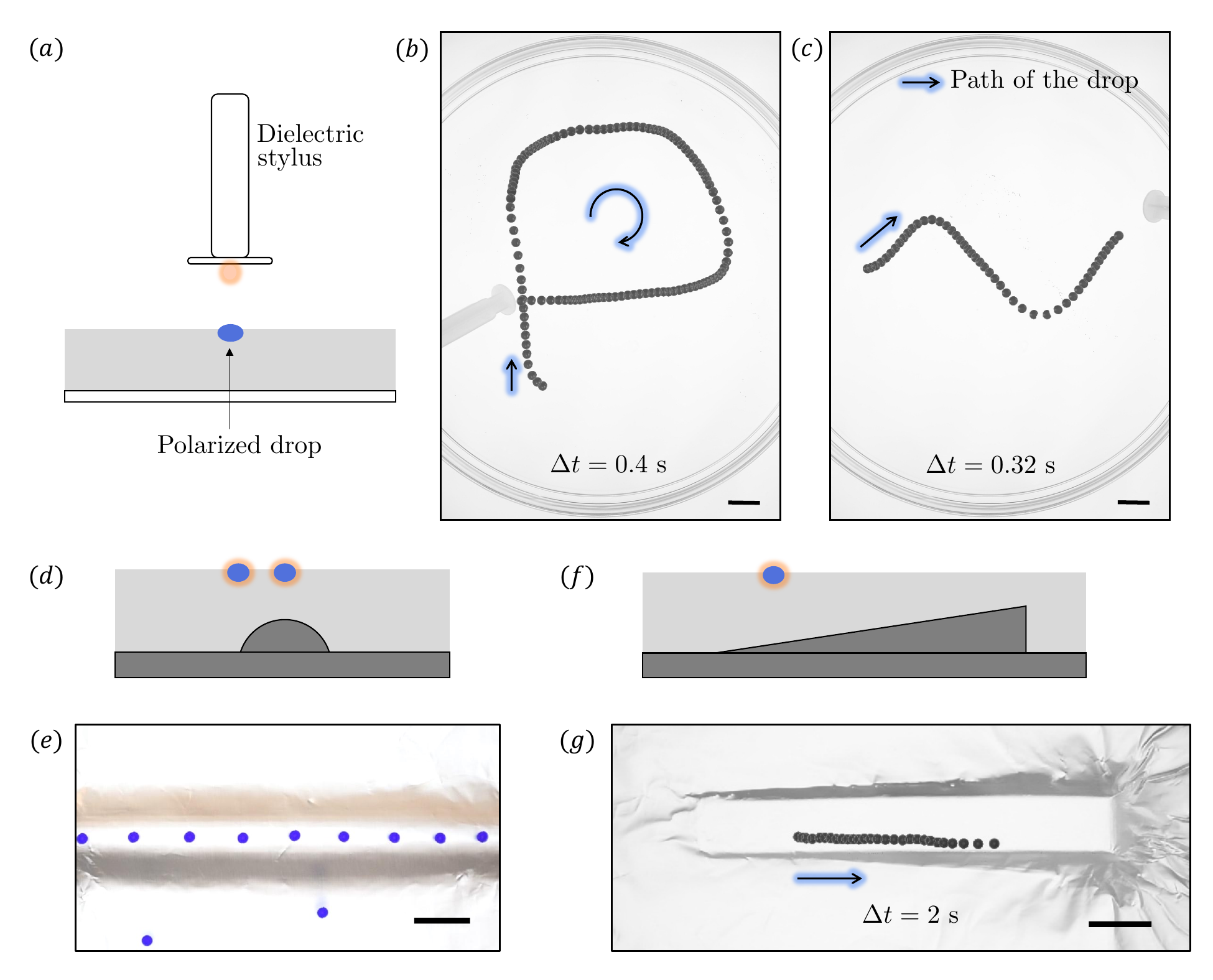}
\caption{\label{ch4_h-patterning} Manipulation of the drops at the interface. (a) Schematic showing configuration 1 with polarized but uncharged drops where the polarization comes from static charges on the surface of a dielectric stylus. (b) An image showing the spatiotemporal evolution of a drop as a stylus is used to write out the letter ``P" with the drop. The time elapsed was $45$ seconds and the arrows with the blue halo show the path of the drop. (c) An image showing another spatiotemporal evolution of a drop as a stylus is used to make a sine wave. The time elapsed was $15$ seconds. (d) Schematic showing a cross-section of the hemispherical cylinder placed on the bottom to create a non-uniform surface. (e) Image showing the drops preferentially assembled and aligned over the ridge. The horizontal feature on the image correspond to the elevated bottom surface. (f) A schematic of the cross-section of the geometry considered in an experiment, where an inclined plane with an angle of about $2^{\circ}$ is placed below the drop. (g) A spatiotemporal image showing the trajectory of the spontaneous motion of a charged drop over the inclined plane. The time elapsed was 48 seconds. All scale bars in the figure represent 1 cm.
}
\end{center}
\end{figure*}

\section{Manipulation of drops}

With the understanding of the collective behavior of the drops in both the polarized and the charged cases, we now offer two ways to manipulate the drops at the oil-air interface.

\subsection{Active control with a stylus}

We showed that static charges present on the surface of the container, which is underneath the drops, can affect the position of the drops at the oil-air interface. We now present a more dynamic, yet simple, approach using a dielectric stylus, as shown schematically in Fig. \ref{ch4_h-patterning}(a). We use the plunger of a syringe as the dielectric stylus and, after neutralizing the surface with a Zerostat pistol, place charges on the flat bottom surface of the stylus using  contact electrification. The stylus can now be used to move the drop as desired on the interface. In Fig. \ref{ch4_h-patterning}(b,c), we provide images showing (b) the spatiotemporal evolution of a single drop as the stylus is used to draw the letter ``P" and (c) an approximate sine wave traced by a single drop. Since the electrostatic interactions are instantaneous and the friction is low due to the low viscosity of the oil subphase, the drops can be manipulated rather quickly. We believe that this method has the potential for future applications due to its simplicity and effectiveness.

\subsection{Passive control with a prescribed shape of the container}
We established that the height of the oil bath $h$ affects the electrostatic potential at the interface. Therefore, patterning the height profile of the bottom boundary creates a non-uniform potential that allows for preferential assembly and spontaneous motion of the drops. Figure \ref{ch4_h-patterning}(d) shows a schematic of the experimental system, where we use a section of a cylinder to create an uneven height profile. Next, Fig. \ref{ch4_h-patterning}(e) shows the experimental image at a quasi-equilibrium state, where the charged drops have moved and arranged themselves along the shallow ridge over the elevated surface. The underlying physics that drives the drops to the shallow region can be explained by the classic problem in electrostatics of a point charge next to a conducting sphere. If a point charge is confined to a horizontal plane of height $h$ above a conducting sphere, the location at which the force along the plane vanishes will be directly above the center of the sphere. Although the ideas remain the same, we consider a hemispherical boss shape to theoretically explain this phenomenon in the Supplementary Information (see SI Sec. V and Fig. S5). 

The feature that the charged drop will move along the interface to be close the conducting bottom surface can be further leveraged to create unmediated motion of the drops along a desired path. In experiments, we placed a charged drop above an inclined plane, as schematically shown in Fig. \ref{ch4_h-patterning}(f), with an inclination angle of about $2^{\circ}$. The drop spontaneously moved toward the shallow end. A spatiotemporal plot showing the trajectory of the drop is displayed in
Fig.~\ref{ch4_h-patterning}(g), where the time between each frame shown is $2$ s. Although the dynamics of the motion will depend on the details of the geometry, as shown in the model in the SI Sec. V, this is a simple way to produce unmediated and spontaneous motion at the interface.

\section{Conclusions}
In conclusion, a novel experimental system was introduced to control and manipulate water drops floating at an oil-air interface. We showed that uncharged drops can be trapped at desired locations on the interface, or effortlessly moved around in determined trajectories along the interface through polarization via static surface charges. On the other hand, a collection of charged drops self-assembled at the interface, dominated by the electrostatic repulsive forces that was found to be a function of the height $h$ of the oil bath. Exploiting the dependence on $h$, we showed that patterning the height profile of the bottom boundary allowed for preferential assembly and directed motion of the drops. With this work, we extend the research area of drop manipulation to the liquid-air interface and also offer our experimental setup as a potential platform to study many other problems in physics including 2D solids, colloidal crystals, and elastic waves.

\acknowledgements{We thank P.-T. Brun, S. Datta, and L. Deike for helpful discussions. JPdS acknowledges support from the Princeton Bioengineering Institute- Innovators (PBI$^2$) Postdoctoral Fellowship. GT is supported by NSF through the Princeton University (PCCM) Materials Research Science and Engineering Center DMR-2011750.}

\appendix

\section{Methods}

All the experiments were done using Fluorinert FC-40 (Sigma-Aldrich) and deionized water dyed with Erioglaucine disodium salt (0.4 wt.\%; Sigma-Aldrich). Since the dye is a salt, a high concentration of the dye ensured that the drops were only modestly charged for the second configuration \cite{sun2015using}. Note that the dye is not necessary for the qualitative observations described in this work. 

In the first experimental configuration, water drops are dispensed continuously from a syringe through a metal needle, which ensured neutral charge of the drops. The neutral charge of the drops were checked using a Faraday cup. In the second configuration, the drops are positively charged. A two-inlet one-outlet microfluidic T-junction device was used to create a segmented flow of water drops in oil \cite{garstecki2006formation}. The flow rate of both phases was $Q=0.05$ mL/min as set using a syringe pump. At the outlet, the solution flowed through a teflon tube, where the water drops picked up a net positive charge due to the charge separation that occurs in the electrical double layer. The amount of charge that the water drops pick up is a function of the length of the tube, diameter of the tube, flow rate, concentration of dye in the drop, size of the drop, and the humidity of the environment. For simplicity, we kept all these parameters constant in the experiments presented here. We note that the experimental system is very sensitive to the external fields present in the environment in the sense that the position and the movement of the drops at the interface can be influenced by fields created by static charges on nearby surfaces including the hand of the experimentalist. 

The container used in the experiments was either a standard petri dish (Fisherbrand; diameter 150 mm), as is or covered with aluminum foil to create a conductive surface, or a stainless steel baking pan also covered with aluminum foil (for the deep bath experiments; $h=0.8,~1.0,~1.1$ cm). The conductive surfaces were electrically grounded to an outlet on the wall. To pattern the height profile of the bottom boundary, we 3D printed (Formlabs form 2) the necessary shapes and placed them on the petri dish before covering with aluminum foil. All experiments were imaged using a Nikon D7500 camera. 


\begin{thebibliography}{41}%
\makeatletter
\providecommand \@ifxundefined [1]{%
 \@ifx{#1\undefined}
}%
\providecommand \@ifnum [1]{%
 \ifnum #1\expandafter \@firstoftwo
 \else \expandafter \@secondoftwo
 \fi
}%
\providecommand \@ifx [1]{%
 \ifx #1\expandafter \@firstoftwo
 \else \expandafter \@secondoftwo
 \fi
}%
\providecommand \natexlab [1]{#1}%
\providecommand \enquote  [1]{``#1''}%
\providecommand \bibnamefont  [1]{#1}%
\providecommand \bibfnamefont [1]{#1}%
\providecommand \citenamefont [1]{#1}%
\providecommand \href@noop [0]{\@secondoftwo}%
\providecommand \href [0]{\begingroup \@sanitize@url \@href}%
\providecommand \@href[1]{\@@startlink{#1}\@@href}%
\providecommand \@@href[1]{\endgroup#1\@@endlink}%
\providecommand \@sanitize@url [0]{\catcode `\\12\catcode `\$12\catcode
  `\&12\catcode `\#12\catcode `\^12\catcode `\_12\catcode `\%12\relax}%
\providecommand \@@startlink[1]{}%
\providecommand \@@endlink[0]{}%
\providecommand \url  [0]{\begingroup\@sanitize@url \@url }%
\providecommand \@url [1]{\endgroup\@href {#1}{\urlprefix }}%
\providecommand \urlprefix  [0]{URL }%
\providecommand \Eprint [0]{\href }%
\providecommand \doibase [0]{https://doi.org/}%
\providecommand \selectlanguage [0]{\@gobble}%
\providecommand \bibinfo  [0]{\@secondoftwo}%
\providecommand \bibfield  [0]{\@secondoftwo}%
\providecommand \translation [1]{[#1]}%
\providecommand \BibitemOpen [0]{}%
\providecommand \bibitemStop [0]{}%
\providecommand \bibitemNoStop [0]{.\EOS\space}%
\providecommand \EOS [0]{\spacefactor3000\relax}%
\providecommand \BibitemShut  [1]{\csname bibitem#1\endcsname}%
\let\auto@bib@innerbib\@empty
\bibitem [{\citenamefont {Konietzko}\ and\ \citenamefont
  {DiMartino}(2013)}]{konietzko2013avatar}%
  \BibitemOpen
  \bibfield  {author} {\bibinfo {author} {\bibfnamefont {B.}~\bibnamefont
  {Konietzko}}\ and\ \bibinfo {author} {\bibfnamefont {M.~D.}\ \bibnamefont
  {DiMartino}},\ }\href@noop {} {\emph {\bibinfo {title} {Avatar: The Last
  Airbender-The Art of the Animated Series}}}\ (\bibinfo  {publisher} {Dark
  Horse Comics},\ \bibinfo {year} {2013})\BibitemShut {NoStop}%
\bibitem [{\citenamefont {Kuhn}\ and\ \citenamefont
  {Myers}(1979)}]{kuhn1979ink}%
  \BibitemOpen
  \bibfield  {author} {\bibinfo {author} {\bibfnamefont {L.}~\bibnamefont
  {Kuhn}}\ and\ \bibinfo {author} {\bibfnamefont {R.~A.}\ \bibnamefont
  {Myers}},\ }\bibfield  {title} {\bibinfo {title} {Ink-jet printing},\
  }\href@noop {} {\bibfield  {journal} {\bibinfo  {journal} {Scientific
  American}\ }\textbf {\bibinfo {volume} {240}},\ \bibinfo {pages} {162}
  (\bibinfo {year} {1979})}\BibitemShut {NoStop}%
\bibitem [{\citenamefont {Basaran}\ \emph {et~al.}(2013)\citenamefont
  {Basaran}, \citenamefont {Gao},\ and\ \citenamefont
  {Bhat}}]{basaran2013nonstandard}%
  \BibitemOpen
  \bibfield  {author} {\bibinfo {author} {\bibfnamefont {O.~A.}\ \bibnamefont
  {Basaran}}, \bibinfo {author} {\bibfnamefont {H.}~\bibnamefont {Gao}},\ and\
  \bibinfo {author} {\bibfnamefont {P.~P.}\ \bibnamefont {Bhat}},\ }\bibfield
  {title} {\bibinfo {title} {Nonstandard inkjets},\ }\href@noop {} {\bibfield
  {journal} {\bibinfo  {journal} {Annual Review of Fluid Mechanics}\ }\textbf
  {\bibinfo {volume} {45}},\ \bibinfo {pages} {85} (\bibinfo {year}
  {2013})}\BibitemShut {NoStop}%
\bibitem [{\citenamefont {Barbulovic-Nad}\ \emph {et~al.}(2008)\citenamefont
  {Barbulovic-Nad}, \citenamefont {Yang}, \citenamefont {Park},\ and\
  \citenamefont {Wheeler}}]{barbulovic2008digital}%
  \BibitemOpen
  \bibfield  {author} {\bibinfo {author} {\bibfnamefont {I.}~\bibnamefont
  {Barbulovic-Nad}}, \bibinfo {author} {\bibfnamefont {H.}~\bibnamefont
  {Yang}}, \bibinfo {author} {\bibfnamefont {P.~S.}\ \bibnamefont {Park}},\
  and\ \bibinfo {author} {\bibfnamefont {A.~R.}\ \bibnamefont {Wheeler}},\
  }\bibfield  {title} {\bibinfo {title} {Digital microfluidics for cell-based
  assays},\ }\href@noop {} {\bibfield  {journal} {\bibinfo  {journal} {Lab on a
  Chip}\ }\textbf {\bibinfo {volume} {8}},\ \bibinfo {pages} {519} (\bibinfo
  {year} {2008})}\BibitemShut {NoStop}%
\bibitem [{\citenamefont {Hajji}\ \emph {et~al.}(2020)\citenamefont {Hajji},
  \citenamefont {Serra}, \citenamefont {Geremie}, \citenamefont {Ferrante},
  \citenamefont {Renault}, \citenamefont {Viovy}, \citenamefont {Descroix},\
  and\ \citenamefont {Ferraro}}]{hajji2020droplet}%
  \BibitemOpen
  \bibfield  {author} {\bibinfo {author} {\bibfnamefont {I.}~\bibnamefont
  {Hajji}}, \bibinfo {author} {\bibfnamefont {M.}~\bibnamefont {Serra}},
  \bibinfo {author} {\bibfnamefont {L.}~\bibnamefont {Geremie}}, \bibinfo
  {author} {\bibfnamefont {I.}~\bibnamefont {Ferrante}}, \bibinfo {author}
  {\bibfnamefont {R.}~\bibnamefont {Renault}}, \bibinfo {author} {\bibfnamefont
  {J.-L.}\ \bibnamefont {Viovy}}, \bibinfo {author} {\bibfnamefont
  {S.}~\bibnamefont {Descroix}},\ and\ \bibinfo {author} {\bibfnamefont
  {D.}~\bibnamefont {Ferraro}},\ }\bibfield  {title} {\bibinfo {title} {Droplet
  microfluidic platform for fast and continuous-flow rt-qpcr analysis devoted
  to cancer diagnosis application},\ }\href@noop {} {\bibfield  {journal}
  {\bibinfo  {journal} {Sensors and Actuators B: Chemical}\ }\textbf {\bibinfo
  {volume} {303}},\ \bibinfo {pages} {127171} (\bibinfo {year}
  {2020})}\BibitemShut {NoStop}%
\bibitem [{\citenamefont {Whitesides}\ \emph {et~al.}(2001)\citenamefont
  {Whitesides}, \citenamefont {Ostuni}, \citenamefont {Takayama}, \citenamefont
  {Jiang},\ and\ \citenamefont {Ingber}}]{whitesides2001soft}%
  \BibitemOpen
  \bibfield  {author} {\bibinfo {author} {\bibfnamefont {G.~M.}\ \bibnamefont
  {Whitesides}}, \bibinfo {author} {\bibfnamefont {E.}~\bibnamefont {Ostuni}},
  \bibinfo {author} {\bibfnamefont {S.}~\bibnamefont {Takayama}}, \bibinfo
  {author} {\bibfnamefont {X.}~\bibnamefont {Jiang}},\ and\ \bibinfo {author}
  {\bibfnamefont {D.~E.}\ \bibnamefont {Ingber}},\ }\bibfield  {title}
  {\bibinfo {title} {Soft lithography in biology and biochemistry},\
  }\href@noop {} {\bibfield  {journal} {\bibinfo  {journal} {Annual Review of
  Biomedical Engineering}\ }\textbf {\bibinfo {volume} {3}},\ \bibinfo {pages}
  {335} (\bibinfo {year} {2001})}\BibitemShut {NoStop}%
\bibitem [{\citenamefont {Sun}\ \emph {et~al.}(2020)\citenamefont {Sun},
  \citenamefont {Steyer}, \citenamefont {Allen}, \citenamefont {Payne},
  \citenamefont {Kennedy},\ and\ \citenamefont {Stephenson}}]{sun2020droplet}%
  \BibitemOpen
  \bibfield  {author} {\bibinfo {author} {\bibfnamefont {A.~C.}\ \bibnamefont
  {Sun}}, \bibinfo {author} {\bibfnamefont {D.~J.}\ \bibnamefont {Steyer}},
  \bibinfo {author} {\bibfnamefont {A.~R.}\ \bibnamefont {Allen}}, \bibinfo
  {author} {\bibfnamefont {E.~M.}\ \bibnamefont {Payne}}, \bibinfo {author}
  {\bibfnamefont {R.~T.}\ \bibnamefont {Kennedy}},\ and\ \bibinfo {author}
  {\bibfnamefont {C.~R.}\ \bibnamefont {Stephenson}},\ }\bibfield  {title}
  {\bibinfo {title} {A droplet microfluidic platform for high-throughput
  photochemical reaction discovery},\ }\href@noop {} {\bibfield  {journal}
  {\bibinfo  {journal} {Nature Communications}\ }\textbf {\bibinfo {volume}
  {11}},\ \bibinfo {pages} {6202} (\bibinfo {year} {2020})}\BibitemShut
  {NoStop}%
\bibitem [{\citenamefont {Li}\ \emph {et~al.}(2020)\citenamefont {Li},
  \citenamefont {Fang}, \citenamefont {Zhao}, \citenamefont {Li}, \citenamefont
  {Li}, \citenamefont {Li}, \citenamefont {Li}, \citenamefont {Feng},\ and\
  \citenamefont {Song}}]{li2020droplet}%
  \BibitemOpen
  \bibfield  {author} {\bibinfo {author} {\bibfnamefont {H.}~\bibnamefont
  {Li}}, \bibinfo {author} {\bibfnamefont {W.}~\bibnamefont {Fang}}, \bibinfo
  {author} {\bibfnamefont {Z.}~\bibnamefont {Zhao}}, \bibinfo {author}
  {\bibfnamefont {A.}~\bibnamefont {Li}}, \bibinfo {author} {\bibfnamefont
  {Z.}~\bibnamefont {Li}}, \bibinfo {author} {\bibfnamefont {M.}~\bibnamefont
  {Li}}, \bibinfo {author} {\bibfnamefont {Q.}~\bibnamefont {Li}}, \bibinfo
  {author} {\bibfnamefont {X.}~\bibnamefont {Feng}},\ and\ \bibinfo {author}
  {\bibfnamefont {Y.}~\bibnamefont {Song}},\ }\bibfield  {title} {\bibinfo
  {title} {Droplet precise self-splitting on patterned adhesive surfaces for
  simultaneous multidetection},\ }\href@noop {} {\bibfield  {journal} {\bibinfo
   {journal} {Angewandte Chemie International Edition}\ }\textbf {\bibinfo
  {volume} {59}},\ \bibinfo {pages} {10535} (\bibinfo {year}
  {2020})}\BibitemShut {NoStop}%
\bibitem [{\citenamefont {Batchelder}(1983)}]{batchelder1983dielectrophoretic}%
  \BibitemOpen
  \bibfield  {author} {\bibinfo {author} {\bibfnamefont {J.}~\bibnamefont
  {Batchelder}},\ }\bibfield  {title} {\bibinfo {title} {Dielectrophoretic
  manipulator},\ }\href@noop {} {\bibfield  {journal} {\bibinfo  {journal}
  {Review of Scientific Instruments}\ }\textbf {\bibinfo {volume} {54}},\
  \bibinfo {pages} {300} (\bibinfo {year} {1983})}\BibitemShut {NoStop}%
\bibitem [{\citenamefont {Jones}(2002)}]{jones2002relationship}%
  \BibitemOpen
  \bibfield  {author} {\bibinfo {author} {\bibfnamefont {T.~B.}\ \bibnamefont
  {Jones}},\ }\bibfield  {title} {\bibinfo {title} {On the relationship of
  dielectrophoresis and electrowetting},\ }\href@noop {} {\bibfield  {journal}
  {\bibinfo  {journal} {Langmuir}\ }\textbf {\bibinfo {volume} {18}},\ \bibinfo
  {pages} {4437} (\bibinfo {year} {2002})}\BibitemShut {NoStop}%
\bibitem [{\citenamefont {Gascoyne}\ \emph {et~al.}(2004)\citenamefont
  {Gascoyne}, \citenamefont {Vykoukal}, \citenamefont {Schwartz}, \citenamefont
  {Anderson}, \citenamefont {Vykoukal}, \citenamefont {Current}, \citenamefont
  {McConaghy}, \citenamefont {Becker},\ and\ \citenamefont
  {Andrews}}]{gascoyne2004dielectrophoresis}%
  \BibitemOpen
  \bibfield  {author} {\bibinfo {author} {\bibfnamefont {P.~R.}\ \bibnamefont
  {Gascoyne}}, \bibinfo {author} {\bibfnamefont {J.~V.}\ \bibnamefont
  {Vykoukal}}, \bibinfo {author} {\bibfnamefont {J.~A.}\ \bibnamefont
  {Schwartz}}, \bibinfo {author} {\bibfnamefont {T.~J.}\ \bibnamefont
  {Anderson}}, \bibinfo {author} {\bibfnamefont {D.~M.}\ \bibnamefont
  {Vykoukal}}, \bibinfo {author} {\bibfnamefont {K.~W.}\ \bibnamefont
  {Current}}, \bibinfo {author} {\bibfnamefont {C.}~\bibnamefont {McConaghy}},
  \bibinfo {author} {\bibfnamefont {F.~F.}\ \bibnamefont {Becker}},\ and\
  \bibinfo {author} {\bibfnamefont {C.}~\bibnamefont {Andrews}},\ }\bibfield
  {title} {\bibinfo {title} {Dielectrophoresis-based programmable fluidic
  processors},\ }\href@noop {} {\bibfield  {journal} {\bibinfo  {journal} {Lab
  on a Chip}\ }\textbf {\bibinfo {volume} {4}},\ \bibinfo {pages} {299}
  (\bibinfo {year} {2004})}\BibitemShut {NoStop}%
\bibitem [{\citenamefont {Hunt}\ \emph {et~al.}(2008)\citenamefont {Hunt},
  \citenamefont {Issadore},\ and\ \citenamefont
  {Westervelt}}]{hunt2008integrated}%
  \BibitemOpen
  \bibfield  {author} {\bibinfo {author} {\bibfnamefont {T.~P.}\ \bibnamefont
  {Hunt}}, \bibinfo {author} {\bibfnamefont {D.}~\bibnamefont {Issadore}},\
  and\ \bibinfo {author} {\bibfnamefont {R.~M.}\ \bibnamefont {Westervelt}},\
  }\bibfield  {title} {\bibinfo {title} {Integrated circuit/microfluidic chip
  to programmably trap and move cells and droplets with dielectrophoresis},\
  }\href@noop {} {\bibfield  {journal} {\bibinfo  {journal} {Lab on a Chip}\
  }\textbf {\bibinfo {volume} {8}},\ \bibinfo {pages} {81} (\bibinfo {year}
  {2008})}\BibitemShut {NoStop}%
\bibitem [{\citenamefont {Beni}\ and\ \citenamefont
  {Hackwood}(1981)}]{beni1981electro}%
  \BibitemOpen
  \bibfield  {author} {\bibinfo {author} {\bibfnamefont {G.}~\bibnamefont
  {Beni}}\ and\ \bibinfo {author} {\bibfnamefont {S.}~\bibnamefont
  {Hackwood}},\ }\bibfield  {title} {\bibinfo {title} {Electro-wetting
  displays},\ }\href@noop {} {\bibfield  {journal} {\bibinfo  {journal}
  {Applied Physics Letters}\ }\textbf {\bibinfo {volume} {38}},\ \bibinfo
  {pages} {207} (\bibinfo {year} {1981})}\BibitemShut {NoStop}%
\bibitem [{\citenamefont {Vallet}\ \emph {et~al.}(1996)\citenamefont {Vallet},
  \citenamefont {Berge},\ and\ \citenamefont
  {Vovelle}}]{vallet1996electrowetting}%
  \BibitemOpen
  \bibfield  {author} {\bibinfo {author} {\bibfnamefont {M.}~\bibnamefont
  {Vallet}}, \bibinfo {author} {\bibfnamefont {B.}~\bibnamefont {Berge}},\ and\
  \bibinfo {author} {\bibfnamefont {L.}~\bibnamefont {Vovelle}},\ }\bibfield
  {title} {\bibinfo {title} {Electrowetting of water and aqueous solutions on
  poly (ethylene terephthalate) insulating films},\ }\href@noop {} {\bibfield
  {journal} {\bibinfo  {journal} {Polymer}\ }\textbf {\bibinfo {volume} {37}},\
  \bibinfo {pages} {2465} (\bibinfo {year} {1996})}\BibitemShut {NoStop}%
\bibitem [{\citenamefont {Pollack}\ \emph {et~al.}(2000)\citenamefont
  {Pollack}, \citenamefont {Fair},\ and\ \citenamefont
  {Shenderov}}]{pollack2000electrowetting}%
  \BibitemOpen
  \bibfield  {author} {\bibinfo {author} {\bibfnamefont {M.~G.}\ \bibnamefont
  {Pollack}}, \bibinfo {author} {\bibfnamefont {R.~B.}\ \bibnamefont {Fair}},\
  and\ \bibinfo {author} {\bibfnamefont {A.~D.}\ \bibnamefont {Shenderov}},\
  }\bibfield  {title} {\bibinfo {title} {Electrowetting-based actuation of
  liquid droplets for microfluidic applications},\ }\href@noop {} {\bibfield
  {journal} {\bibinfo  {journal} {Applied Physics Letters}\ }\textbf {\bibinfo
  {volume} {77}},\ \bibinfo {pages} {1725} (\bibinfo {year}
  {2000})}\BibitemShut {NoStop}%
\bibitem [{\citenamefont {Lee}\ \emph {et~al.}(2002)\citenamefont {Lee},
  \citenamefont {Moon}, \citenamefont {Fowler}, \citenamefont {Schoellhammer},\
  and\ \citenamefont {Kim}}]{lee2002electrowetting}%
  \BibitemOpen
  \bibfield  {author} {\bibinfo {author} {\bibfnamefont {J.}~\bibnamefont
  {Lee}}, \bibinfo {author} {\bibfnamefont {H.}~\bibnamefont {Moon}}, \bibinfo
  {author} {\bibfnamefont {J.}~\bibnamefont {Fowler}}, \bibinfo {author}
  {\bibfnamefont {T.}~\bibnamefont {Schoellhammer}},\ and\ \bibinfo {author}
  {\bibfnamefont {C.-J.}\ \bibnamefont {Kim}},\ }\bibfield  {title} {\bibinfo
  {title} {Electrowetting and electrowetting-on-dielectric for microscale
  liquid handling},\ }\href@noop {} {\bibfield  {journal} {\bibinfo  {journal}
  {Sensors and Actuators A: Physical}\ }\textbf {\bibinfo {volume} {95}},\
  \bibinfo {pages} {259} (\bibinfo {year} {2002})}\BibitemShut {NoStop}%
\bibitem [{\citenamefont {Nelson}\ and\ \citenamefont
  {Kim}(2012)}]{nelson2012droplet}%
  \BibitemOpen
  \bibfield  {author} {\bibinfo {author} {\bibfnamefont {W.~C.}\ \bibnamefont
  {Nelson}}\ and\ \bibinfo {author} {\bibfnamefont {C.-J.}\ \bibnamefont
  {Kim}},\ }\bibfield  {title} {\bibinfo {title} {Droplet actuation by
  electrowetting-on-dielectric (ewod): A review},\ }\href@noop {} {\bibfield
  {journal} {\bibinfo  {journal} {Journal of Adhesion Science and Technology}\
  }\textbf {\bibinfo {volume} {26}},\ \bibinfo {pages} {1747} (\bibinfo {year}
  {2012})}\BibitemShut {NoStop}%
\bibitem [{\citenamefont {Peng}\ \emph {et~al.}(2014)\citenamefont {Peng},
  \citenamefont {Zhang}, \citenamefont {Ju} \emph {et~al.}}]{peng2014ewod}%
  \BibitemOpen
  \bibfield  {author} {\bibinfo {author} {\bibfnamefont {C.}~\bibnamefont
  {Peng}}, \bibinfo {author} {\bibfnamefont {Z.}~\bibnamefont {Zhang}},
  \bibinfo {author} {\bibfnamefont {Y.~S.}\ \bibnamefont {Ju}}, \emph
  {et~al.},\ }\bibfield  {title} {\bibinfo {title} {Ewod (electrowetting on
  dielectric) digital microfluidics powered by finger actuation},\ }\href@noop
  {} {\bibfield  {journal} {\bibinfo  {journal} {Lab on a Chip}\ }\textbf
  {\bibinfo {volume} {14}},\ \bibinfo {pages} {1117} (\bibinfo {year}
  {2014})}\BibitemShut {NoStop}%
\bibitem [{\citenamefont {Li}\ \emph {et~al.}(2019)\citenamefont {Li},
  \citenamefont {Ha}, \citenamefont {Liu}, \citenamefont {van Dam},\ and\
  \citenamefont {Kim}}]{li2019ionic}%
  \BibitemOpen
  \bibfield  {author} {\bibinfo {author} {\bibfnamefont {J.}~\bibnamefont
  {Li}}, \bibinfo {author} {\bibfnamefont {N.~S.}\ \bibnamefont {Ha}}, \bibinfo
  {author} {\bibfnamefont {T.}~\bibnamefont {Liu}}, \bibinfo {author}
  {\bibfnamefont {R.~M.}\ \bibnamefont {van Dam}},\ and\ \bibinfo {author}
  {\bibfnamefont {C.-J.}\ \bibnamefont {Kim}},\ }\bibfield  {title} {\bibinfo
  {title} {Ionic-surfactant-mediated electro-dewetting for digital
  microfluidics},\ }\href@noop {} {\bibfield  {journal} {\bibinfo  {journal}
  {Nature}\ }\textbf {\bibinfo {volume} {572}},\ \bibinfo {pages} {507}
  (\bibinfo {year} {2019})}\BibitemShut {NoStop}%
\bibitem [{\citenamefont {Jin}\ \emph {et~al.}(2022)\citenamefont {Jin},
  \citenamefont {Xu}, \citenamefont {Zhang}, \citenamefont {Li}, \citenamefont
  {Sun}, \citenamefont {Yang}, \citenamefont {Liu}, \citenamefont {Mao},\ and\
  \citenamefont {Wang}}]{jin2022electrostatic}%
  \BibitemOpen
  \bibfield  {author} {\bibinfo {author} {\bibfnamefont {Y.}~\bibnamefont
  {Jin}}, \bibinfo {author} {\bibfnamefont {W.}~\bibnamefont {Xu}}, \bibinfo
  {author} {\bibfnamefont {H.}~\bibnamefont {Zhang}}, \bibinfo {author}
  {\bibfnamefont {R.}~\bibnamefont {Li}}, \bibinfo {author} {\bibfnamefont
  {J.}~\bibnamefont {Sun}}, \bibinfo {author} {\bibfnamefont {S.}~\bibnamefont
  {Yang}}, \bibinfo {author} {\bibfnamefont {M.}~\bibnamefont {Liu}}, \bibinfo
  {author} {\bibfnamefont {H.}~\bibnamefont {Mao}},\ and\ \bibinfo {author}
  {\bibfnamefont {Z.}~\bibnamefont {Wang}},\ }\bibfield  {title} {\bibinfo
  {title} {Electrostatic tweezer for droplet manipulation},\ }\href@noop {}
  {\bibfield  {journal} {\bibinfo  {journal} {Proceedings of the National
  Academy of Sciences}\ }\textbf {\bibinfo {volume} {119}},\ \bibinfo {pages}
  {e2105459119} (\bibinfo {year} {2022})}\BibitemShut {NoStop}%
\bibitem [{\citenamefont {Xu}\ \emph {et~al.}(2022)\citenamefont {Xu},
  \citenamefont {Jin}, \citenamefont {Li}, \citenamefont {Song}, \citenamefont
  {Gao}, \citenamefont {Zhang}, \citenamefont {Wang}, \citenamefont {Cui},
  \citenamefont {Yan},\ and\ \citenamefont {Wang}}]{xu2022triboelectric}%
  \BibitemOpen
  \bibfield  {author} {\bibinfo {author} {\bibfnamefont {W.}~\bibnamefont
  {Xu}}, \bibinfo {author} {\bibfnamefont {Y.}~\bibnamefont {Jin}}, \bibinfo
  {author} {\bibfnamefont {W.}~\bibnamefont {Li}}, \bibinfo {author}
  {\bibfnamefont {Y.}~\bibnamefont {Song}}, \bibinfo {author} {\bibfnamefont
  {S.}~\bibnamefont {Gao}}, \bibinfo {author} {\bibfnamefont {B.}~\bibnamefont
  {Zhang}}, \bibinfo {author} {\bibfnamefont {L.}~\bibnamefont {Wang}},
  \bibinfo {author} {\bibfnamefont {M.}~\bibnamefont {Cui}}, \bibinfo {author}
  {\bibfnamefont {X.}~\bibnamefont {Yan}},\ and\ \bibinfo {author}
  {\bibfnamefont {Z.}~\bibnamefont {Wang}},\ }\bibfield  {title} {\bibinfo
  {title} {Triboelectric wetting for continuous droplet transport},\
  }\href@noop {} {\bibfield  {journal} {\bibinfo  {journal} {Science Advances}\
  }\textbf {\bibinfo {volume} {8}},\ \bibinfo {pages} {eade2085} (\bibinfo
  {year} {2022})}\BibitemShut {NoStop}%
\bibitem [{\citenamefont {Jin}\ \emph {et~al.}(2023)\citenamefont {Jin},
  \citenamefont {Liu}, \citenamefont {Xu}, \citenamefont {Sun}, \citenamefont
  {Huang}, \citenamefont {Yang}, \citenamefont {Yang}, \citenamefont {Wang},
  \citenamefont {Lam}, \citenamefont {Li} \emph {et~al.}}]{jin2023charge}%
  \BibitemOpen
  \bibfield  {author} {\bibinfo {author} {\bibfnamefont {Y.}~\bibnamefont
  {Jin}}, \bibinfo {author} {\bibfnamefont {X.}~\bibnamefont {Liu}}, \bibinfo
  {author} {\bibfnamefont {W.}~\bibnamefont {Xu}}, \bibinfo {author}
  {\bibfnamefont {P.}~\bibnamefont {Sun}}, \bibinfo {author} {\bibfnamefont
  {S.}~\bibnamefont {Huang}}, \bibinfo {author} {\bibfnamefont
  {S.}~\bibnamefont {Yang}}, \bibinfo {author} {\bibfnamefont {X.}~\bibnamefont
  {Yang}}, \bibinfo {author} {\bibfnamefont {Q.}~\bibnamefont {Wang}}, \bibinfo
  {author} {\bibfnamefont {R.~H.}\ \bibnamefont {Lam}}, \bibinfo {author}
  {\bibfnamefont {R.}~\bibnamefont {Li}}, \emph {et~al.},\ }\bibfield  {title}
  {\bibinfo {title} {Charge-powered electrotaxis for versatile droplet
  manipulation},\ }\href@noop {} {\bibfield  {journal} {\bibinfo  {journal}
  {ACS Nano}\ } (\bibinfo {year} {2023})}\BibitemShut {NoStop}%
\bibitem [{\citenamefont {Zhang}\ \emph {et~al.}(2022)\citenamefont {Zhang},
  \citenamefont {Jiang}, \citenamefont {Hu}, \citenamefont {Wu}, \citenamefont
  {Zhang}, \citenamefont {Li}, \citenamefont {Li}, \citenamefont {Zhang},
  \citenamefont {Wu}, \citenamefont {Ding} \emph
  {et~al.}}]{zhang2022reconfigurable}%
  \BibitemOpen
  \bibfield  {author} {\bibinfo {author} {\bibfnamefont {Y.}~\bibnamefont
  {Zhang}}, \bibinfo {author} {\bibfnamefont {S.}~\bibnamefont {Jiang}},
  \bibinfo {author} {\bibfnamefont {Y.}~\bibnamefont {Hu}}, \bibinfo {author}
  {\bibfnamefont {T.}~\bibnamefont {Wu}}, \bibinfo {author} {\bibfnamefont
  {Y.}~\bibnamefont {Zhang}}, \bibinfo {author} {\bibfnamefont
  {H.}~\bibnamefont {Li}}, \bibinfo {author} {\bibfnamefont {A.}~\bibnamefont
  {Li}}, \bibinfo {author} {\bibfnamefont {Y.}~\bibnamefont {Zhang}}, \bibinfo
  {author} {\bibfnamefont {H.}~\bibnamefont {Wu}}, \bibinfo {author}
  {\bibfnamefont {Y.}~\bibnamefont {Ding}}, \emph {et~al.},\ }\bibfield
  {title} {\bibinfo {title} {Reconfigurable magnetic liquid metal robot for
  high-performance droplet manipulation},\ }\href@noop {} {\bibfield  {journal}
  {\bibinfo  {journal} {Nano Letters}\ }\textbf {\bibinfo {volume} {22}},\
  \bibinfo {pages} {2923} (\bibinfo {year} {2022})}\BibitemShut {NoStop}%
\bibitem [{\citenamefont {Xu}\ \emph {et~al.}(2023)\citenamefont {Xu},
  \citenamefont {Yao}, \citenamefont {Deng}, \citenamefont {Fang},
  \citenamefont {Dupont}, \citenamefont {Zhang}, \citenamefont {{\v{C}}opar},
  \citenamefont {Tkalec},\ and\ \citenamefont
  {Wang}}]{xu2023magnetocontrollable}%
  \BibitemOpen
  \bibfield  {author} {\bibinfo {author} {\bibfnamefont {Y.}~\bibnamefont
  {Xu}}, \bibinfo {author} {\bibfnamefont {Y.}~\bibnamefont {Yao}}, \bibinfo
  {author} {\bibfnamefont {W.}~\bibnamefont {Deng}}, \bibinfo {author}
  {\bibfnamefont {J.-C.}\ \bibnamefont {Fang}}, \bibinfo {author}
  {\bibfnamefont {R.~L.}\ \bibnamefont {Dupont}}, \bibinfo {author}
  {\bibfnamefont {M.}~\bibnamefont {Zhang}}, \bibinfo {author} {\bibfnamefont
  {S.}~\bibnamefont {{\v{C}}opar}}, \bibinfo {author} {\bibfnamefont
  {U.}~\bibnamefont {Tkalec}},\ and\ \bibinfo {author} {\bibfnamefont
  {X.}~\bibnamefont {Wang}},\ }\bibfield  {title} {\bibinfo {title}
  {Magnetocontrollable droplet mobility on liquid crystal-infused porous
  surfaces},\ }\href@noop {} {\bibfield  {journal} {\bibinfo  {journal} {Nano
  Research}\ }\textbf {\bibinfo {volume} {16}},\ \bibinfo {pages} {5098}
  (\bibinfo {year} {2023})}\BibitemShut {NoStop}%
\bibitem [{\citenamefont {Liu}\ \emph {et~al.}(2022)\citenamefont {Liu},
  \citenamefont {Li}, \citenamefont {Peng}, \citenamefont {Chen},\ and\
  \citenamefont {Zhang}}]{liu2022simple}%
  \BibitemOpen
  \bibfield  {author} {\bibinfo {author} {\bibfnamefont {M.}~\bibnamefont
  {Liu}}, \bibinfo {author} {\bibfnamefont {C.}~\bibnamefont {Li}}, \bibinfo
  {author} {\bibfnamefont {Z.}~\bibnamefont {Peng}}, \bibinfo {author}
  {\bibfnamefont {S.}~\bibnamefont {Chen}},\ and\ \bibinfo {author}
  {\bibfnamefont {B.}~\bibnamefont {Zhang}},\ }\bibfield  {title} {\bibinfo
  {title} {Simple but efficient method to transport droplets on arbitrarily
  controllable paths},\ }\href@noop {} {\bibfield  {journal} {\bibinfo
  {journal} {Langmuir}\ }\textbf {\bibinfo {volume} {38}},\ \bibinfo {pages}
  {3917} (\bibinfo {year} {2022})}\BibitemShut {NoStop}%
\bibitem [{\citenamefont {Yuan}\ \emph {et~al.}(2023)\citenamefont {Yuan},
  \citenamefont {Lu}, \citenamefont {Liu}, \citenamefont {Bai}, \citenamefont
  {Zhao}, \citenamefont {Feng},\ and\ \citenamefont
  {Liu}}]{yuan2023ultrasonic}%
  \BibitemOpen
  \bibfield  {author} {\bibinfo {author} {\bibfnamefont {Z.}~\bibnamefont
  {Yuan}}, \bibinfo {author} {\bibfnamefont {C.}~\bibnamefont {Lu}}, \bibinfo
  {author} {\bibfnamefont {C.}~\bibnamefont {Liu}}, \bibinfo {author}
  {\bibfnamefont {X.}~\bibnamefont {Bai}}, \bibinfo {author} {\bibfnamefont
  {L.}~\bibnamefont {Zhao}}, \bibinfo {author} {\bibfnamefont {S.}~\bibnamefont
  {Feng}},\ and\ \bibinfo {author} {\bibfnamefont {Y.}~\bibnamefont {Liu}},\
  }\bibfield  {title} {\bibinfo {title} {Ultrasonic tweezer for multifunctional
  droplet manipulation},\ }\href@noop {} {\bibfield  {journal} {\bibinfo
  {journal} {Science Advances}\ }\textbf {\bibinfo {volume} {9}},\ \bibinfo
  {pages} {eadg2352} (\bibinfo {year} {2023})}\BibitemShut {NoStop}%
\bibitem [{\citenamefont {Luo}\ \emph {et~al.}(2023)\citenamefont {Luo},
  \citenamefont {Liu}, \citenamefont {Zhou}, \citenamefont {Zhang},
  \citenamefont {Chen}, \citenamefont {Zhan}, \citenamefont {Hu}, \citenamefont
  {He}, \citenamefont {Xie}, \citenamefont {Huan} \emph
  {et~al.}}]{luo2023contactless}%
  \BibitemOpen
  \bibfield  {author} {\bibinfo {author} {\bibfnamefont {T.}~\bibnamefont
  {Luo}}, \bibinfo {author} {\bibfnamefont {S.}~\bibnamefont {Liu}}, \bibinfo
  {author} {\bibfnamefont {R.}~\bibnamefont {Zhou}}, \bibinfo {author}
  {\bibfnamefont {C.}~\bibnamefont {Zhang}}, \bibinfo {author} {\bibfnamefont
  {D.}~\bibnamefont {Chen}}, \bibinfo {author} {\bibfnamefont {Y.}~\bibnamefont
  {Zhan}}, \bibinfo {author} {\bibfnamefont {Q.}~\bibnamefont {Hu}}, \bibinfo
  {author} {\bibfnamefont {X.}~\bibnamefont {He}}, \bibinfo {author}
  {\bibfnamefont {Y.}~\bibnamefont {Xie}}, \bibinfo {author} {\bibfnamefont
  {Z.}~\bibnamefont {Huan}}, \emph {et~al.},\ }\bibfield  {title} {\bibinfo
  {title} {Contactless acoustic tweezer for droplet manipulation on
  superhydrophobic surfaces},\ }\href@noop {} {\bibfield  {journal} {\bibinfo
  {journal} {Lab on a Chip}\ }\textbf {\bibinfo {volume} {23}},\ \bibinfo
  {pages} {3989} (\bibinfo {year} {2023})}\BibitemShut {NoStop}%
\bibitem [{\citenamefont {Wang}\ \emph {et~al.}(2022)\citenamefont {Wang},
  \citenamefont {Liu}, \citenamefont {Liu}, \citenamefont {Zhao}, \citenamefont
  {Wang}, \citenamefont {Wang},\ and\ \citenamefont {Du}}]{wang2022light}%
  \BibitemOpen
  \bibfield  {author} {\bibinfo {author} {\bibfnamefont {F.}~\bibnamefont
  {Wang}}, \bibinfo {author} {\bibfnamefont {M.}~\bibnamefont {Liu}}, \bibinfo
  {author} {\bibfnamefont {C.}~\bibnamefont {Liu}}, \bibinfo {author}
  {\bibfnamefont {Q.}~\bibnamefont {Zhao}}, \bibinfo {author} {\bibfnamefont
  {T.}~\bibnamefont {Wang}}, \bibinfo {author} {\bibfnamefont {Z.}~\bibnamefont
  {Wang}},\ and\ \bibinfo {author} {\bibfnamefont {X.}~\bibnamefont {Du}},\
  }\bibfield  {title} {\bibinfo {title} {Light-induced charged slippery
  surfaces},\ }\href@noop {} {\bibfield  {journal} {\bibinfo  {journal}
  {Science Advances}\ }\textbf {\bibinfo {volume} {8}},\ \bibinfo {pages}
  {eabp9369} (\bibinfo {year} {2022})}\BibitemShut {NoStop}%
\bibitem [{\citenamefont {Dong}\ \emph {et~al.}(2023)\citenamefont {Dong},
  \citenamefont {Zhou}, \citenamefont {Tang}, \citenamefont {Chen},\ and\
  \citenamefont {Huang}}]{dong2023laser}%
  \BibitemOpen
  \bibfield  {author} {\bibinfo {author} {\bibfnamefont {J.}~\bibnamefont
  {Dong}}, \bibinfo {author} {\bibfnamefont {J.}~\bibnamefont {Zhou}}, \bibinfo
  {author} {\bibfnamefont {H.}~\bibnamefont {Tang}}, \bibinfo {author}
  {\bibfnamefont {B.}~\bibnamefont {Chen}},\ and\ \bibinfo {author}
  {\bibfnamefont {L.}~\bibnamefont {Huang}},\ }\bibfield  {title} {\bibinfo
  {title} {Laser-guided programmable construction of cell-laden hydrogel
  microstructures for in vitro drug evaluation},\ }\href@noop {} {\bibfield
  {journal} {\bibinfo  {journal} {Biofabrication}\ }\textbf {\bibinfo {volume}
  {15}},\ \bibinfo {pages} {045011} (\bibinfo {year} {2023})}\BibitemShut
  {NoStop}%
\bibitem [{\citenamefont {Gifford}\ and\ \citenamefont
  {Scriven}(1971)}]{gifford1971attraction}%
  \BibitemOpen
  \bibfield  {author} {\bibinfo {author} {\bibfnamefont {W.}~\bibnamefont
  {Gifford}}\ and\ \bibinfo {author} {\bibfnamefont {L.}~\bibnamefont
  {Scriven}},\ }\bibfield  {title} {\bibinfo {title} {On the attraction of
  floating particles},\ }\href@noop {} {\bibfield  {journal} {\bibinfo
  {journal} {Chemical Engineering Science}\ }\textbf {\bibinfo {volume} {26}},\
  \bibinfo {pages} {287} (\bibinfo {year} {1971})}\BibitemShut {NoStop}%
\bibitem [{\citenamefont {Vella}\ and\ \citenamefont
  {Mahadevan}(2005)}]{vella2005cheerios}%
  \BibitemOpen
  \bibfield  {author} {\bibinfo {author} {\bibfnamefont {D.}~\bibnamefont
  {Vella}}\ and\ \bibinfo {author} {\bibfnamefont {L.}~\bibnamefont
  {Mahadevan}},\ }\bibfield  {title} {\bibinfo {title} {The “cheerios
  effect”},\ }\href@noop {} {\bibfield  {journal} {\bibinfo  {journal}
  {American Journal of Physics}\ }\textbf {\bibinfo {volume} {73}},\ \bibinfo
  {pages} {817} (\bibinfo {year} {2005})}\BibitemShut {NoStop}%
\bibitem [{\citenamefont {Couder}\ \emph {et~al.}(2005)\citenamefont {Couder},
  \citenamefont {Protiere}, \citenamefont {Fort},\ and\ \citenamefont
  {Boudaoud}}]{couder2005walking}%
  \BibitemOpen
  \bibfield  {author} {\bibinfo {author} {\bibfnamefont {Y.}~\bibnamefont
  {Couder}}, \bibinfo {author} {\bibfnamefont {S.}~\bibnamefont {Protiere}},
  \bibinfo {author} {\bibfnamefont {E.}~\bibnamefont {Fort}},\ and\ \bibinfo
  {author} {\bibfnamefont {A.}~\bibnamefont {Boudaoud}},\ }\bibfield  {title}
  {\bibinfo {title} {Walking and orbiting droplets},\ }\href@noop {} {\bibfield
   {journal} {\bibinfo  {journal} {Nature}\ }\textbf {\bibinfo {volume}
  {437}},\ \bibinfo {pages} {208} (\bibinfo {year} {2005})}\BibitemShut
  {NoStop}%
\bibitem [{\citenamefont {Protiere}\ \emph {et~al.}(2006)\citenamefont
  {Protiere}, \citenamefont {Boudaoud},\ and\ \citenamefont
  {Couder}}]{protiere2006particle}%
  \BibitemOpen
  \bibfield  {author} {\bibinfo {author} {\bibfnamefont {S.}~\bibnamefont
  {Protiere}}, \bibinfo {author} {\bibfnamefont {A.}~\bibnamefont {Boudaoud}},\
  and\ \bibinfo {author} {\bibfnamefont {Y.}~\bibnamefont {Couder}},\
  }\bibfield  {title} {\bibinfo {title} {Particle--wave association on a fluid
  interface},\ }\href@noop {} {\bibfield  {journal} {\bibinfo  {journal}
  {Journal of Fluid Mechanics}\ }\textbf {\bibinfo {volume} {554}},\ \bibinfo
  {pages} {85} (\bibinfo {year} {2006})}\BibitemShut {NoStop}%
\bibitem [{\citenamefont {Eddi}\ \emph {et~al.}(2008)\citenamefont {Eddi},
  \citenamefont {Terwagne}, \citenamefont {Fort},\ and\ \citenamefont
  {Couder}}]{eddi2008wave}%
  \BibitemOpen
  \bibfield  {author} {\bibinfo {author} {\bibfnamefont {A.}~\bibnamefont
  {Eddi}}, \bibinfo {author} {\bibfnamefont {D.}~\bibnamefont {Terwagne}},
  \bibinfo {author} {\bibfnamefont {E.}~\bibnamefont {Fort}},\ and\ \bibinfo
  {author} {\bibfnamefont {Y.}~\bibnamefont {Couder}},\ }\bibfield  {title}
  {\bibinfo {title} {Wave propelled ratchets and drifting rafts},\ }\href@noop
  {} {\bibfield  {journal} {\bibinfo  {journal} {Europhysics Letters}\ }\textbf
  {\bibinfo {volume} {82}},\ \bibinfo {pages} {44001} (\bibinfo {year}
  {2008})}\BibitemShut {NoStop}%
\bibitem [{\citenamefont {Couchman}\ and\ \citenamefont
  {Bush}(2020)}]{couchman2020free}%
  \BibitemOpen
  \bibfield  {author} {\bibinfo {author} {\bibfnamefont {M.~M.}\ \bibnamefont
  {Couchman}}\ and\ \bibinfo {author} {\bibfnamefont {J.~W.}\ \bibnamefont
  {Bush}},\ }\bibfield  {title} {\bibinfo {title} {Free rings of bouncing
  droplets: stability and dynamics},\ }\href@noop {} {\bibfield  {journal}
  {\bibinfo  {journal} {Journal of Fluid Mechanics}\ }\textbf {\bibinfo
  {volume} {903}},\ \bibinfo {pages} {A49} (\bibinfo {year}
  {2020})}\BibitemShut {NoStop}%
\bibitem [{\citenamefont {Majhy}\ \emph {et~al.}(2020)\citenamefont {Majhy},
  \citenamefont {Jain},\ and\ \citenamefont {Sen}}]{majhy2020attraction}%
  \BibitemOpen
  \bibfield  {author} {\bibinfo {author} {\bibfnamefont {B.}~\bibnamefont
  {Majhy}}, \bibinfo {author} {\bibfnamefont {S.}~\bibnamefont {Jain}},\ and\
  \bibinfo {author} {\bibfnamefont {A.}~\bibnamefont {Sen}},\ }\bibfield
  {title} {\bibinfo {title} {Attraction and repulsion between liquid droplets
  over a liquid-impregnated surface},\ }\href@noop {} {\bibfield  {journal}
  {\bibinfo  {journal} {The Journal of Physical Chemistry Letters}\ }\textbf
  {\bibinfo {volume} {11}},\ \bibinfo {pages} {10001} (\bibinfo {year}
  {2020})}\BibitemShut {NoStop}%
\bibitem [{\citenamefont {Li}\ \emph {et~al.}(2023)\citenamefont {Li},
  \citenamefont {Pahlavan}, \citenamefont {Chen}, \citenamefont {Liu},
  \citenamefont {Li}, \citenamefont {Stone},\ and\ \citenamefont
  {Granick}}]{li2023oil}%
  \BibitemOpen
  \bibfield  {author} {\bibinfo {author} {\bibfnamefont {Y.}~\bibnamefont
  {Li}}, \bibinfo {author} {\bibfnamefont {A.~A.}\ \bibnamefont {Pahlavan}},
  \bibinfo {author} {\bibfnamefont {Y.}~\bibnamefont {Chen}}, \bibinfo {author}
  {\bibfnamefont {S.}~\bibnamefont {Liu}}, \bibinfo {author} {\bibfnamefont
  {Y.}~\bibnamefont {Li}}, \bibinfo {author} {\bibfnamefont {H.~A.}\
  \bibnamefont {Stone}},\ and\ \bibinfo {author} {\bibfnamefont
  {S.}~\bibnamefont {Granick}},\ }\bibfield  {title} {\bibinfo {title}
  {Oil-on-water droplets faceted and stabilized by vortex halos in the
  subphase},\ }\href@noop {} {\bibfield  {journal} {\bibinfo  {journal}
  {Proceedings of the National Academy of Sciences}\ }\textbf {\bibinfo
  {volume} {120}},\ \bibinfo {pages} {e2214657120} (\bibinfo {year}
  {2023})}\BibitemShut {NoStop}%
\bibitem [{\citenamefont {McCarty}\ and\ \citenamefont
  {Whitesides}(2008)}]{mccarty2008electrostatic}%
  \BibitemOpen
  \bibfield  {author} {\bibinfo {author} {\bibfnamefont {L.~S.}\ \bibnamefont
  {McCarty}}\ and\ \bibinfo {author} {\bibfnamefont {G.~M.}\ \bibnamefont
  {Whitesides}},\ }\bibfield  {title} {\bibinfo {title} {Electrostatic charging
  due to separation of ions at interfaces: contact electrification of ionic
  electrets},\ }\href@noop {} {\bibfield  {journal} {\bibinfo  {journal}
  {Angewandte Chemie International Edition}\ }\textbf {\bibinfo {volume}
  {47}},\ \bibinfo {pages} {2188} (\bibinfo {year} {2008})}\BibitemShut
  {NoStop}%
\bibitem [{\citenamefont {Sun}\ \emph {et~al.}(2015)\citenamefont {Sun},
  \citenamefont {Huang},\ and\ \citenamefont {Soh}}]{sun2015using}%
  \BibitemOpen
  \bibfield  {author} {\bibinfo {author} {\bibfnamefont {Y.}~\bibnamefont
  {Sun}}, \bibinfo {author} {\bibfnamefont {X.}~\bibnamefont {Huang}},\ and\
  \bibinfo {author} {\bibfnamefont {S.}~\bibnamefont {Soh}},\ }\bibfield
  {title} {\bibinfo {title} {Using the gravitational energy of water to
  generate power by separation of charge at interfaces},\ }\href@noop {}
  {\bibfield  {journal} {\bibinfo  {journal} {Chemical Science}\ }\textbf
  {\bibinfo {volume} {6}},\ \bibinfo {pages} {3347} (\bibinfo {year}
  {2015})}\BibitemShut {NoStop}%
\bibitem [{\citenamefont {Choi}\ \emph {et~al.}(2013)\citenamefont {Choi},
  \citenamefont {Lee}, \citenamefont {Im}, \citenamefont {Kang}, \citenamefont
  {Lim}, \citenamefont {Kim},\ and\ \citenamefont
  {Kang}}]{choi2013spontaneous}%
  \BibitemOpen
  \bibfield  {author} {\bibinfo {author} {\bibfnamefont {D.}~\bibnamefont
  {Choi}}, \bibinfo {author} {\bibfnamefont {H.}~\bibnamefont {Lee}}, \bibinfo
  {author} {\bibfnamefont {D.~J.}\ \bibnamefont {Im}}, \bibinfo {author}
  {\bibfnamefont {I.~S.}\ \bibnamefont {Kang}}, \bibinfo {author}
  {\bibfnamefont {G.}~\bibnamefont {Lim}}, \bibinfo {author} {\bibfnamefont
  {D.~S.}\ \bibnamefont {Kim}},\ and\ \bibinfo {author} {\bibfnamefont {K.~H.}\
  \bibnamefont {Kang}},\ }\bibfield  {title} {\bibinfo {title} {Spontaneous
  electrical charging of droplets by conventional pipetting},\ }\href@noop {}
  {\bibfield  {journal} {\bibinfo  {journal} {Scientific Reports}\ }\textbf
  {\bibinfo {volume} {3}},\ \bibinfo {pages} {2037} (\bibinfo {year}
  {2013})}\BibitemShut {NoStop}%
\bibitem [{\citenamefont {Garstecki}\ \emph {et~al.}(2006)\citenamefont
  {Garstecki}, \citenamefont {Fuerstman}, \citenamefont {Stone},\ and\
  \citenamefont {Whitesides}}]{garstecki2006formation}%
  \BibitemOpen
  \bibfield  {author} {\bibinfo {author} {\bibfnamefont {P.}~\bibnamefont
  {Garstecki}}, \bibinfo {author} {\bibfnamefont {M.~J.}\ \bibnamefont
  {Fuerstman}}, \bibinfo {author} {\bibfnamefont {H.~A.}\ \bibnamefont
  {Stone}},\ and\ \bibinfo {author} {\bibfnamefont {G.~M.}\ \bibnamefont
  {Whitesides}},\ }\bibfield  {title} {\bibinfo {title} {Formation of droplets
  and bubbles in a microfluidic t-junction—scaling and mechanism of
  break-up},\ }\href@noop {} {\bibfield  {journal} {\bibinfo  {journal} {Lab on
  a Chip}\ }\textbf {\bibinfo {volume} {6}},\ \bibinfo {pages} {437} (\bibinfo
  {year} {2006})}\BibitemShut {NoStop}%
\end{thebibliography}
\end{document}



\title{Supplementary Information for ``Triboelectrically mediated self-assembly and manipulation of drops at an interface"}

 \author{Paul R. Kaneelil}
    \affiliation{Department of Mechanical and Aerospace Engineering, Princeton University, Princeton, New Jersey 08544, USA}
 \author{J. Pedro de Souza}
    \affiliation{Omenn-Darling Bioengineering Institute, Princeton University, Princeton, New Jersey 08544, USA}
 \author{Günther Turk}
    \affiliation{Princeton Materials Institute, Princeton University, Princeton, New Jersey 08544, USA}
    
 \author{Amir A. Pahlavan}
    \affiliation{Department of Mechanical Engineering and Material Science, Yale University, New Haven, Connecticut 06511, USA}
 \author{Howard A. Stone}
    \affiliation{Department of Mechanical and Aerospace Engineering, Princeton University, Princeton, New Jersey 08544, USA}

\date{\today}

\maketitle

\section{Charge of the drops}
\begin{figure}[b]
\begin{center}
\includegraphics[width=.45\textwidth]{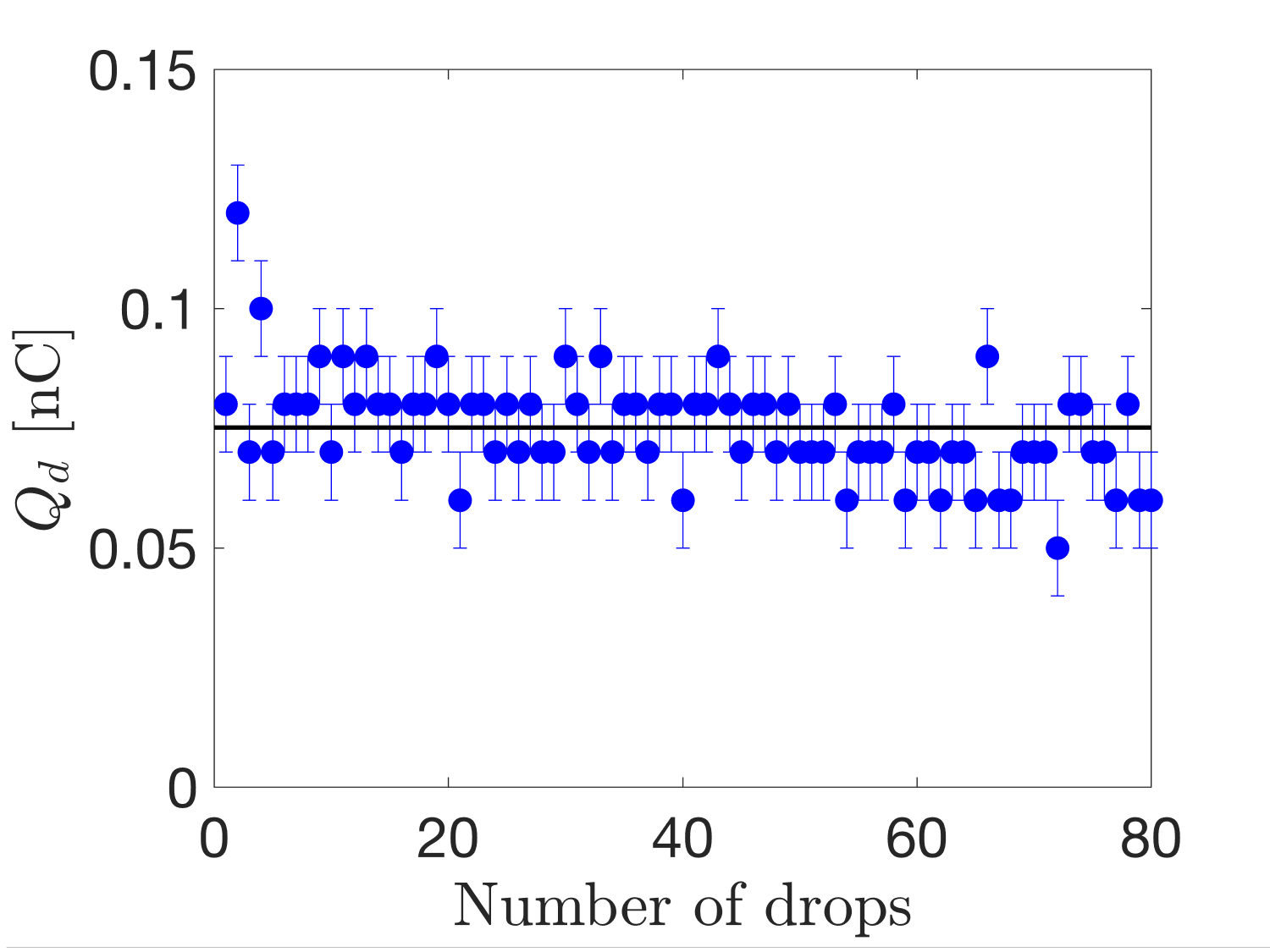}
\caption{\label{charge} The charge of the drops as measured using a Faraday cup and a NanoCoulomb meter. The error bars represent the measurement error and the solid black line represents the average of the data.}
\end{center}
\end{figure}

In all the experiments for configuration 2, we used the same protocol the create the drops to ensure that the drops are uniformly charged (see Methods section). Since the charging relies on contact charge separation, which can be irregular compared to a more controlled method using electrodes and high voltages, we note that the drops may not be charged the exact same amount. We measured the charge of the drops using a Faraday cup and a NanoCoulomb meter (Monroe Electronics; Model 284). Figure \ref{charge} shows the charge of the drops as they were sequentially dispensed into the Faraday cup. The error bars represent the inaccuracy of the measurement. On average, the drops had a charge of $Q_d=0.07 \pm 0.01 $ nC. 

\section{Electrostatic model}\label{secA1}
We derive a simplified theoretical model to elucidate the dependence of the collective behavior of the drops on the bath height $h$. Consider the geometry shown in Fig. 3(e) in the main text, where the drop is assumed to be a flat circle of radius $R$ located at the origin with a uniform charge distribution. We consider a domain that is infinite in the radial direction $r$ and semi-infinite in the axial direction $z$, where there is a finite height $h$ of oil below the drop with a relative permittivity $\epsilon_r$ and an infinite layer of air above the drop. The electrostatic potential $\phi$ for this axisymmetric system satisfies the Laplace equation

\begin{equation}
    \nabla^2\phi = \frac{1}{r}\frac{\partial}{\partial r} \bigg( r \frac{\partial \phi}{\partial r} \bigg ) + \frac{\partial^2 \phi}{\partial z^2}=0.
    \label{eqch4Laplace}
\end{equation}

in both the air phase ($z>0$) and the oil phase ($z<0$). The boundary conditions in the $z$-direction specifies a vanishing electric field far away from the drop, a jump in the field at the interface, and a vanishing potential at the conductive lower boundary:
\begin{subequations}
\label{eqch4_zbc}
\begin{gather}
\frac{\partial \phi}{\partial z}(r, z \rightarrow \infty) = 0, \label{eqch4_zbc1}
\\
\frac{\partial \phi}{\partial z}(r,z=0^+) - \epsilon_r \frac{\partial \phi}{\partial z}(r,z=0^-) = -\frac{q_d}{\epsilon_0}~{\cal H}(R-r), \label{eqch4_zbc2}
\\
\phi(r,z=-h) = 0. \label{eqch4_zbc3}
\end{gather}
\end{subequations}
Here, $\epsilon_0 = 8.854 \times 10^{-12}$ F/m is the permittivity of free space, $q_d$ is the charge per area on the drop, and ${\cal H}$ is the Heaviside step function. The boundary conditions in the $r$-direction specifies symmetry at the center and a vanishing electric field far away:
\begin{subequations}
\label{eqch4_rbc}
\begin{gather}
\frac{\partial \phi}{\partial r}(r = 0, z) = 0, \label{eqch4_rbc1}
\\
\frac{\partial \phi}{\partial r}(r \rightarrow \infty, z) = 0. \label{eqch4_rbc2}
\end{gather}
\end{subequations}

Taking advantage of the circular symmetry and to readily satisfy the boundary conditions in the $r$-direction, we use the Hankel transform to solve Eqs. \ref{eqch4Laplace} - \ref{eqch4_rbc}. The Hankel transform of $\phi$, represented by the transformed variable $\Phi(k,z)$, and the inverse transform are defined as
\begin{subequations}
\label{eqch4hankel}
\begin{gather}
    \Phi(k,z) = \int_{0}^{\infty} \phi(r,z)~J_0(kr)r \,dr, \\
    \phi(r,z) = \int_{0}^{\infty} \Phi(k,z)~J_0(kr)k \,dk
\end{gather}
\end{subequations}
where $J_0$ is the Bessel function of the first kind of order zero and $k$ is the transformed independent variable.

The transformed Eq. \ref{eqch4Laplace} becomes $\frac{\partial^2 \Phi}{\partial z^2} - k^2 \Phi = 0$. After satisfying the boundary conditions, the solution is
\begin{equation}
    \Phi(k,z) = 
        \begin{cases}
      A(k)~e^{-kz},~~~~~~~~~~~~~~ z \geq 0\\
      B(k)~\text{sinh}[k(z+h)],~~z \leq 0\\
    \end{cases} 
    \label{eqch4PhiSol}
\end{equation}
where $A(k) = \frac{Q(k)~\text{sinh}(kh)}{k\epsilon_0[\epsilon_r \text{cosh}(kh) + \text{sinh}(kh)]}$, $B(k) = \frac{Q(k)}{k\epsilon_0[\epsilon_r \text{cosh}(kh) + \text{sinh}(kh)]}$, $Q(k)=\frac{Q_d}{\pi k R}J_1(k R)$, and $Q_d$ is the total charge on the drop.

\begin{figure}[bt]
\begin{center}
\includegraphics[width=0.95\textwidth]{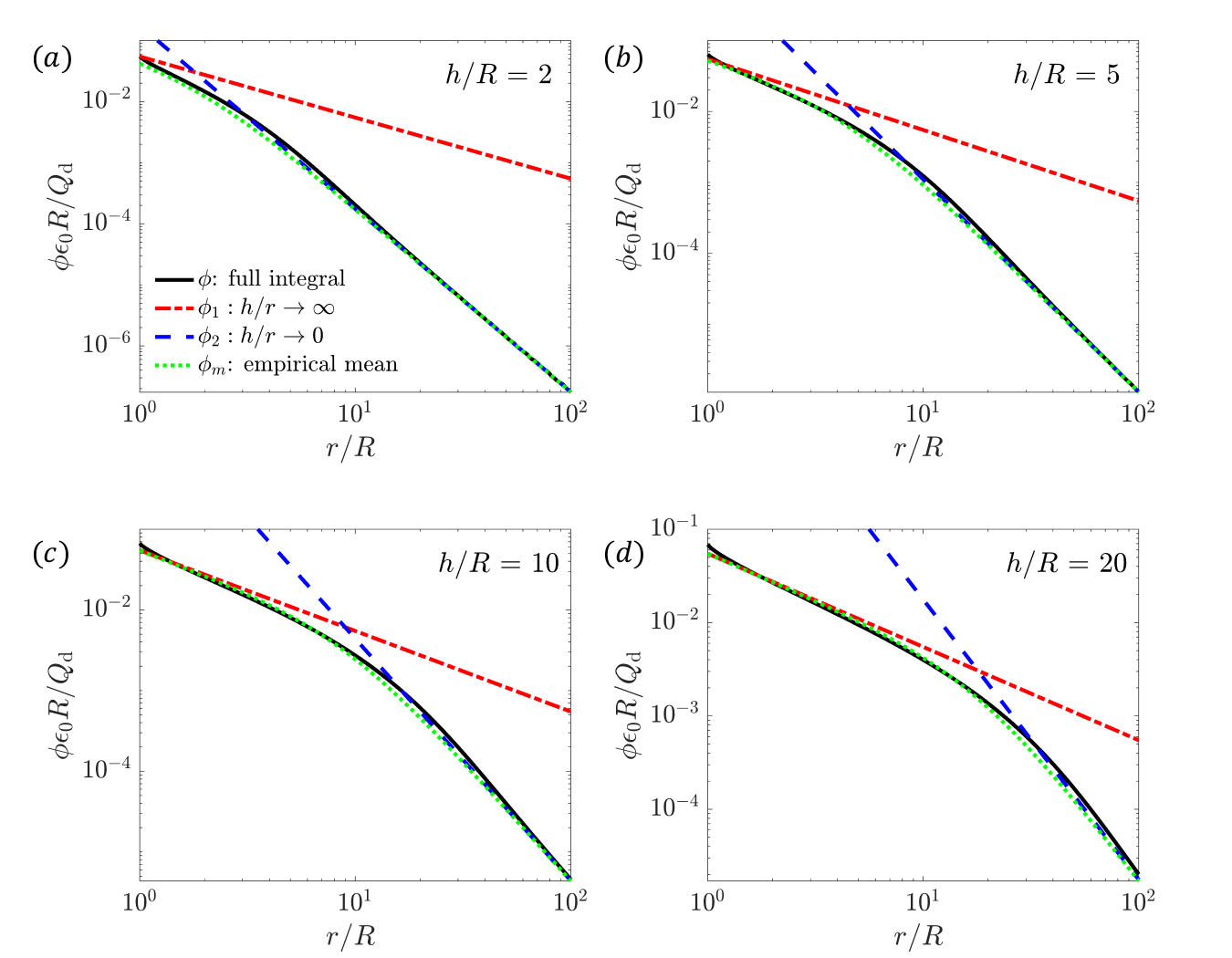}
\caption{\label{appx_theory} The theoretical electric potential near a drop. The black line is the exact solution from the integral in Eq. \ref{eqch4hankel}b, the red dashed-dotted line corresponds to the deep bath limit from Eq. \ref{eqch4deepbath}, the blue dashed line corresponds to the shallow bath limit from Eq. \ref{eqch4shallowbath}, and the green dotted line corresponds to the empirical mean from Eq. 1 in the main text. We plot the solutions for four different values of dimensionless bath height: (a) $h/R=2$, (b) $h/R=5$, (c) $h/R=10$, and (d) $h/R=20$. }
\end{center}
\end{figure}

The inverse transform integral does not have a closed form analytical solution. But, we can approximate the integral analytically and calculate the pair potential at the interface, $z=0$, at two relevant limits. When the bath is deep $kh \rightarrow \infty$, and the drop is small $kR \rightarrow 0$ compared to the separation between two drops, i.e. $h/r \rightarrow \infty$, we get
\begin{equation}
    \phi_1(r,z=0) \approx \frac{Q_d}{4 \pi \epsilon_0 r} \bigg(\frac{2}{\epsilon_r+1} \bigg).
    \label{eqch4deepbath}
\end{equation}
In this limit, the potential is independent of the height of the bath. Conversely, when the height of the bath and the size of the drop are small compared to the separation between two drops, $kh \rightarrow 0$ and $kR \rightarrow 0$, i.e. $h/r \rightarrow 0$, we get
\begin{equation}
    \phi_2(r,z=0) \approx \frac{2 Q_d h^2}{4 \pi \epsilon_0 \epsilon_r^2 r^3}.
    \label{eqch4shallowbath}
\end{equation}
In this limit, the potential has a quadratic dependence on the height of the oil bath. Notice also that the potential has an inverse square dependence on the dielectric constant of the oil bath, which explains why the electrostatic interactions and self-assembly are not seen when using an oil with large $\epsilon_r$. These approximations also assume that the drop is a point charge ($kR \rightarrow 0$). In Fig. \ref{appx_theory}(a)-(d), we plot the exact solution and the two approximations for four different $h/R$ values. Note that non-dimensionalizing the length with $R$ artificially introduces $R$ back into the two approximations. This is done simply to plot all the solutions together which reveals more insight: the deep bath approximation (red dashed-dotted line) agrees with exact solution around the vicinity of the drop ($r/R=1$) as $h/R$ increases, whereas the shallow bath approximation (blue dashed line) agrees with the exact solution farther away from the drop ($r/R\gg 1$) as $h/R$ increases.

We also note that the full potential at the interface $\phi(r,z=0)$ can be approximately represented as the scaled Harmonic mean of the two limits as, $\phi_m \approx 1/2~ H(\phi_1,\phi_2) = (1/\phi_1 + 1/\phi_2)^{-1}$, and is given in Eq. 1 in the main text. While $\phi_m$ (green dotted line) is built on the point charge assumption, we see that it agrees well with the exact solution $\phi$ for finite $R$, as shown in Fig. \ref{appx_theory}. This agreement seems to only hold for small relative permittivity, namely $\epsilon_r<2$. We use the analytical form $\phi_m$ to make quantitative comparison between the theory and the experiments.

\begin{figure}[t]
\begin{center}
\includegraphics[width=0.45\textwidth]{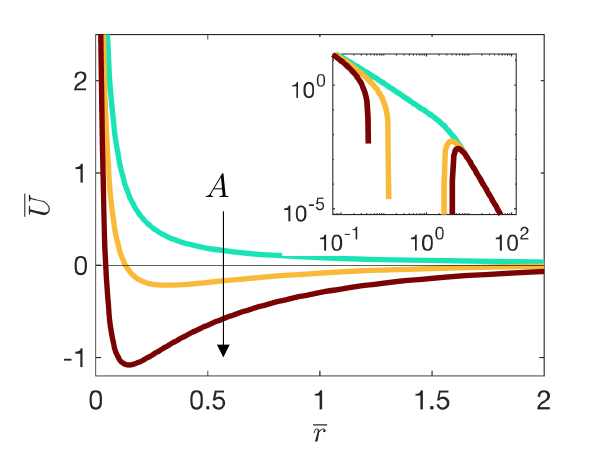}
\caption{\label{si_energy} The dimensionless total energy represented by Eq. \ref{eqch4U_rescale} with $A=0.1,~0.5,$ and $1$ for green, yellow, and brown colored lines, respectively, and $B=7$. }
\end{center}
\end{figure}
\section{Energy of the drops at the interface}

The total energy of a pair of drops interacting at the oil-air interface might include a contribution from the capillary interactions and the electrostatic interactions. Assuming the capillary interactions are similar to that of particles at an interface \cite{majhy2020attraction}, the total energy can be written as follows:
\begin{equation}
    U(r) = -2 \pi \gamma R^2 B^2 \Sigma^2 K_0(r/l_c) + \frac{Q_d^2 h^2}{2 \pi \epsilon_0 r[(\epsilon_r+1)h^2+\epsilon_r^2r^2]},
    \label{eqch4U}
\end{equation}
where $\Sigma=\left(\frac{2D-1}{3}-\frac{1}{2}\text{cos}(\theta)+\frac{1}{6}\text{cos}^3(\theta)\right)$, $D = \rho_w/\rho_o$ is the density ratio of water to oil in this case, $B=\rho_o g R^2/\gamma$ is the bond number, $l_c=\sqrt{\gamma/(\rho_o g)}$ is the capillary length, and $K_0$ is the modified Bessel function of the second kind of order zero. We rescale the equation such that $\overline{U} = U/U_c$ and $\overline{r} = r/l_c$, to get

\begin{equation}
    \overline{U}(\overline{r}) = -A K_0(\overline{r})+\frac{1}{\overline{r}^3+B\overline{r}},
    \label{eqch4U_rescale}
\end{equation}
where $U_c = \frac{Q_d h^2}{2 \pi \epsilon_0 \epsilon_r^2 l_c^3}$, $A = \frac{4\pi^2 \gamma R^2 B^2 \Sigma^2 \epsilon_0 \epsilon_r^2 l_c^3}{Q_d^2 h^2}$, and $B=\frac{(\epsilon_r+1)h^2}{\epsilon_r^2 l_c^2}$. Note that both the capillary and electrostatic interactions are derived based on the assumption that the potentials can be superposed, neglecting non-linear effects. The equilibrium distance $\overline{r}_{eq} = \overline{\ell}_{eq}=\ell_{eq}/l_c$ between the drops correspond to the minimum energy state that must satisfy $\frac{d \overline{U}}{d \overline{r}}=0$. However, we find that this condition does not predict the experimental results. In other words, the total energy in Eq. \ref{eqch4U_rescale} does not have a local minimum, as shown in Fig. \ref{si_energy}, at experimentally relevant values of $A \approx 0.1$ and $B \approx 7$. If $A$ is slightly larger, a local minimum starts to appear but it occurs at $\ell_{eq}/l_c = {\cal O}(0.1)$ or smaller. But, the quasi-equilibrium distances observed in experiments are about $\ell_{f}/l_c \approx 10$. This discrepancy suggests that the interactions between the drops are dominated by electrostatic repulsion and that the system indeed is not in equilibrium within the observation time in experiments. The drops must be continually repelling and moving away from each other.

\begin{figure}[b]
\begin{center}
\includegraphics[width=0.42\textwidth]{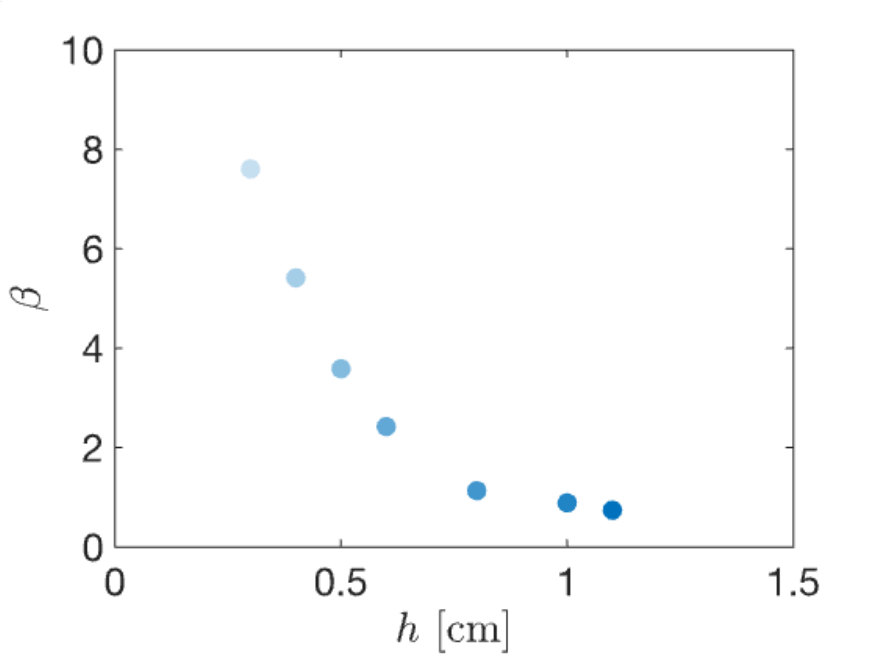}
\caption{\label{si_beta} The fitting parameter $\beta$ as a function of the height of the oil bath $h$. }
\end{center}
\end{figure}
\section{Fitting parameter}
Balancing viscous stresses and electrostatic repulsion, we showed that the theoretical velocity of the charged drop as it repels another can be modelled as $\textbf{V} = -\frac{Q_d}{6 \beta \pi \mu_o R} \frac{d}{dr}[\phi_m(r,z=0)]$, where $\beta$ is a fitting parameter that accounts for any excess drag due to the presence of the interface and the bottom surface of the container in shallow cases. In Fig. \ref{si_beta}, we report the values of $\beta$ used to fit the experimental results to the model in Fig. 3(g) in the main text. The value of $\beta$ was calculated such that the mean of $\langle V/V_{\textrm{exp}}\rangle$ became unity for each data set. We note that $\beta \approx {\cal O}(1)$ in all cases and is larger for smaller values of $h$, as expected, since the point charge approximation breaks down as the size of the drop becomes comparable to the height.

\section{The hemispherical boss}\label{secA3}
\begin{figure}[bt]
\begin{center}
\includegraphics[width=.95\textwidth]{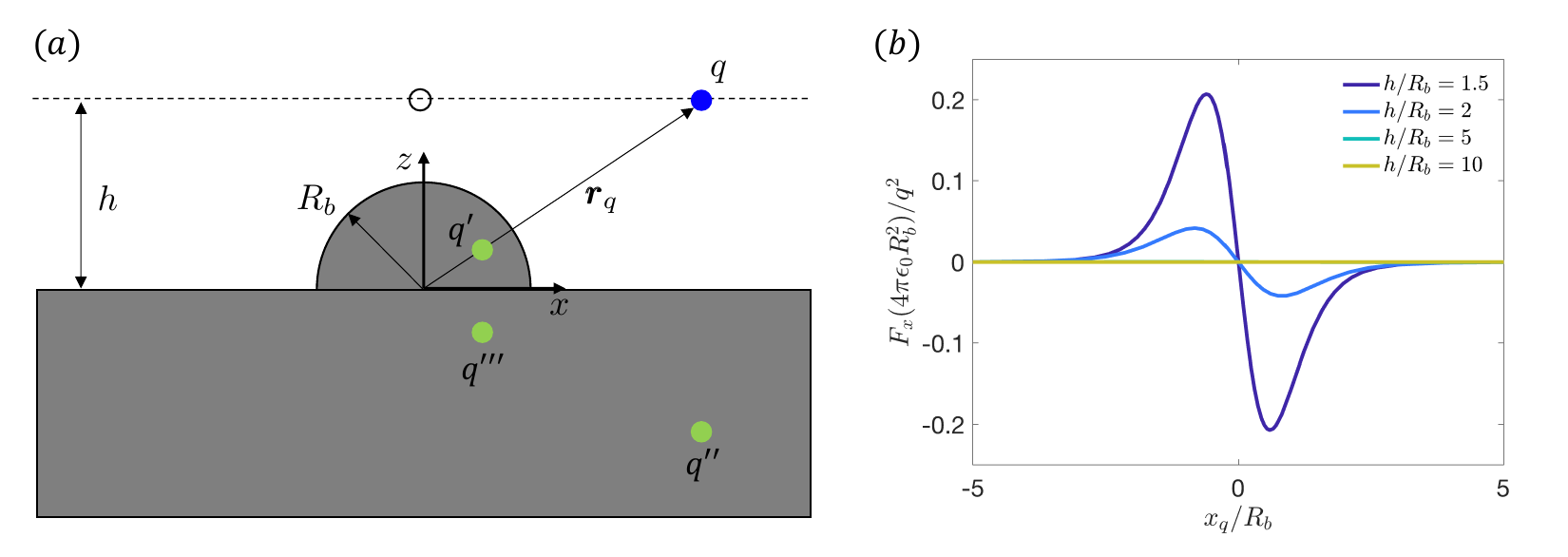}
\caption{\label{ch4_boss} The hemispherical boss. (a) Schematic of the hemispherical boss shape studied in a model, where a point charge $q$ is restricted to be at a height $z=h$. The image charges are shown as green circles. The force-free point directly above the boss, where the distance between the charge $q$ and the conductor becomes minimal, is indicated by an open circle. (b) The $x$-component of the theorerical force on a charge $q$ in the hemispherical boss configuration for various $h/R_b$ values.}
\end{center}
\end{figure}

As discussed in the main text, we analyze the electrostatic effect that drives the drops towards shallow regions above a conductor qualitatively by considering the textbook example of a point charge $q$ near a conducting plane with a hemispherical boss of radius $R_b$ \cite{Jeans_1908}, as shown in Fig. \ref{ch4_boss}(a). Taking into account that the charged drop is confined to the surface of the suspending oil bath, the point charge is restricted to be at a height $z=h$ above the conducting plane. With respect to the center of the hemispherical boss, the position vector of the charge is then $\mathbf{r}_q=(x_q,0,h)^T$. Using a method of images it can be shown that in the $x$-direction a force acts on the charge with the magnitude 

\begin{equation}
\label{boss_xforce}
    F_x = F_x^{\circ}
    + \frac{q^2}{4\pi\epsilon_0}\frac{R_b}{r_q} \left(1-\frac{R_b^2}{r_q^2}\right)
    \left[x_q^2 \left(1-\frac{R_b^2}{r_q^2}\right)^2 + h^2 \left(1+\frac{R_b^2}{r_q^2}\right)^2\right]^{-\frac{3}{2}}\, x_q,
\end{equation}where we have defined the force $F_x^{\circ}$ on a point charge due to a sphere of radius $R_b$ centered at the origin,

\begin{equation}
    F_x^{\circ} = -\frac{q^2}{4\pi\epsilon_0} \frac{R_b}{r_q^4} 
    \left(1-\frac{R_b^2}{r_q^2}\right)^{-2}\,x_q.
\end{equation}It is then easy to see that the charge is forced towards $x_q=0$, where the force in $x$-direction vanishes. This is the point where the distance between the charge and the conductor becomes minimal. Equation \ref{boss_xforce} is plotted in Fig. \ref{ch4_boss}(b) in a non-dimensional form for various non-dimensional height $h/R_b$. Notice that the force vanishes at $x_q/R_b=0$ for small $h/R_b$, and that the force becomes independent of position as $h/R_b$ increases, as expected.  

